\documentclass[prd,aps,showpacs,12pt,amssymb,preprintnumbers,amsmath,amsfonts,nofootinbib]{revtex4-1}

\usepackage{latexsym}
\usepackage[utf8]{inputenc}
\usepackage{amsmath}
\usepackage{cancel}
\usepackage{mathtools}
\usepackage{graphicx}
\usepackage{hyperref}  
\usepackage{dcolumn}                                   
\usepackage{color}
\usepackage{slashed}
\usepackage{ulem}
\usepackage{braket}	

\def\q{\boldsymbol{q}}

\def\k{\boldsymbol{k}}

\def\u{\boldsymbol{u}}

\def\v{\boldsymbol{v}}

\def\D{\mathcal{D}}

\def\D{\mathcal{D}}
\def\La{\mathcal{L}}
\def\d{\partial}
\def\eps{\epsilon}

\newcolumntype{P}[1]{>{\centering\arraybackslash}p{#1}}
\newcolumntype{M}[1]{>{\centering\arraybackslash}m{#1}}
	
\begin{document}
\date{\today}
\title{On-shell effective field theory and \\ quantum transport for hard photons} 

\author{Marc Comadran${}^{1,2,3}$}
\email{mcomadca60@alumnes.ub.edu}

\author{Cristina Manuel${}^{1,2}$}
\email{cristina.manuel@csic.es}

\affiliation{${}^1$Instituto de Ciencias del Espacio (ICE, CSIC) \\
C. Can Magrans s.n., 08193 Cerdanyola del Vall\`es,  Spain 
and 
\\
${}^2$Institut d'Estudis Espacials de Catalunya (IEEC) \\
08860 Castelldefels (Barcelona),  Spain and
}
\affiliation{${}^3$Departament de F\'isica Qu\`antica i Astrof\'isica 
and Institut de Ci\`encies del Cosmos, Universitat de Barcelona (IEEC-UB), Mart\'i i Franqu\'es 1, 08028 Barcelona,  Spain}

\begin{abstract}
We develop an effective field theory for the description of high energetic or hard photons, the on-shell effective theory (OSEFT). 
The OSEFT describes the so called eikonal or semi-classical optical limit, allowing for corrections  organized in  a systematic expansion on inverse  powers of the photon energy.  We derive the OSEFT  from the Maxwell Lagrangian, and study its different properties, such as the gauge symmetry and reparametrization invariance. The theory can be finally formulated in terms of a gauge invariant vector gauge field, without the need to introduce  gauge-fixing.
We then use the OSEFT to  compute corrections to the Wigner photon function, and derive its associated side jump effect from reparametrization invariance. Finally, we discuss how to properly define the Stokes parameters from transport theory once quantum effects are considered, so as to preserve their well-defined properties under Lorentz transformations.
 
\end{abstract}
%\newpage
		
\maketitle
	
\tableofcontents 

%%%%%%%%%%%%%%%%%%%%%%%%%%%%%%%%%%%%%%%%%%%%%%%%%%%%%%%%%%%%%%%%%%%
\section{Introduction}
%%%%%%%%%%%%%%%%%%%%%%%%%%%%%%%%%%%%%%%%%%%%%%%%%%%%%%%%%%%%%%%%%%%

Effective field theories are one of the most useful and advantageous tools in physics. They rely on the idea that in order to describe some phenomena at a given energy scale it is enough to identify the degrees of freedom that operate at that scale, and uncover the Lagrangian that governs their dynamics, exploiting the symmetries of the problem. The effective Lagrangian is then organized in operators of increasing dimension over powers of the high  energy scale. 
A different set of effective field theories have been proposed to describe a wide variety of physical phenomena, see the excellent reviews \cite{Manohar:1996cq,Manohar:2018aog,Becher:2014oda} to discuss the most relevant technical details and most used effective field theories for vacuum physics.

In this manuscript we will focus on the so called on-shell effective field theory (OSEFT) \cite{Manuel:2014dza,Manuel:2016wqs,Manuel:2016cit,Carignano:2018gqt,Carignano:2019zsh,Manuel:2021oah,Comadran:2021pkv}, which was first developed to describe  energetic chiral fermions, with the main aim to describe quantum corrections to the classical transport equations \cite{Manuel:2014dza}. The OSEFT also proved to be useful to compute power  corrections to photon self-energy diagrams in high temperature $T$ plasmas \cite{Manuel:2016wqs}, or mass corrections to the same amplitudes \cite{Comadran:2021pkv}, taking profit of the
natural hierarchy of energy scales that appear in these systems.

Our focus in this manuscript is to describe the OSEFT associated to high energetic or hard photons. This presents different challenges, among them, a proper description respectful with the gauge symmetry of the photon degrees of freedom.
We first present the complete OSEFT associated to an electromagnetic field, as being derived from the Maxwell Lagrangian. As we will discuss, the leading order Lagrangian in a high energy expansion 
describes the so called eikonal or optical limit, while higher order terms in such an expansion would describe quantum corrections to the eikonal limit. Notably, the effective field theory can be formulated without imposing any gauge fixing condition. The non-physical components of 
the vector gauge field potential can be eliminated, after integrating out and using local field redefinitions. Remarkably, we will show that the final OSEFT Lagrangian only contains one vector degree of freedom and demonstrate its invariance under gauge transformations that respect the energy scaling of the effective photon field. By studying the reparametrization invariance (RI) of the theory, we also check that the OSEFT for photons is respectful with the Lorentz symmetry.

Then, we also consider a many-body system, such as a thermal plasma. By employing the OSEFT for photons and well-established thermal field theory techniques, we derive the quantum kinetic equations obeyed by the photon Wigner function. Also, we show that the OSEFT allows us to systematically compute quantum corrections associated to the photon Wigner function.

The polarization space components of the photon Wigner function  can be related to the so called Stokes parameters \cite{jackson}. While the Stokes parameters are not Lorentz invariant, some ratios of them are \cite{Cocke,Han:1997zz}. For example, the percentage of circular polarization of a system is expected to be a Lorentz invariant. In this work, we will see that when quantum corrections are taken into consideration, the classical definition of the polarization ratio is no longer Lorentz invariant. This is related to the fact that in quantum kinetic theory, Lorentz transformations acquire non trivial modifications when quantum effects are taken into account \cite{Chen:2014cla}. The unusual transformation properties have meaningful physical implications, for instance the so called side jump effect \cite{Chen:2015gta,Stone:2015kla}. Hence, we propose how to generalize the definition of the Stokes parameters, so as to preserve the Lorentz invariance of the polarization ratios when small quantum effects are taken into account.
As we will discuss, the modification is relevant when the frame of reference of observation is not at rest with the thermal ensemble where the photon radiation is produced. For instance, this could be realized in many astrophysical and cosmological settings, where those conditions are usually met.

This manuscript is organized as follows, in Sec.\ref{II} we develop the OSEFT associated to highly energetic photons. In Sec.\ref{II.A} we derive the OSEFT vacuum propagator and the dispersion relation, also, in Sec.\ref{II.B} we construct an effective gauge field for almost on-shell photons and in Sec.\ref{II.C} we discuss how the gauge symmetry is realized in the OSEFT. Then, in Sec.\ref{II.D} we introduce a polarization basis in the effective field theory, while Sec.\ref{II.E} is devoted to the study of the RI  of the theory. In Sec.\ref{II.F} we explain the relation between the full theory and OSEFT variables. Then we employ the OSEFT in Sec.\ref{III} to develop a quantum kinetic theory for hard photons. Using those results, in Sec.\ref{III.A} we construct the photon Wigner function while in Sec.\ref{III.B} a derivation of the side jump effect for photons from first principles is presented.  In addition, in Sec.\ref{IV} we discuss the Lorentz transformation properties of the Stokes parameters when small quantum effects are taken into account, from a quantum kinetic theory perspective. We present our conclusions in Sec.\ref{V}. Finally, we elaborate in App.\ref{A} a simplified operator notation used throughout the paper and explain some operator identities used in this work. In App.\ref{B} we give general RI transformations for the effective photons fields which are too lengthy to include in the main text.

We use natural units $c= \hbar= k_B=1$, metric conventions $\text{diag}(g^{\mu\nu})=(1,-1,-1,-1)$ and the normalization $\eps^{0123}=1$ for the Levi-Civita tensor. Sometimes we will use the shortened notations $A^{(\mu}B^{\nu)}=A^\mu B^\nu+A^\nu B^\mu$ and $A^{[\mu}B^{\nu]}=A^\mu B^\nu-A^\nu B^\mu$. We employ boldface letters for three dimensional vectors e.g $A^\mu=(A^0,A^i)=(A^0, \boldsymbol{A})$.

\newpage 

%%%%%%%%%%%%%%%%%%%%%%%%%%%%%%%%%%%%%%%%%%%%%%%%%%%%%%%%%%%%%%%%%%%
\section{On-shell effective field theory for photons}
%%%%%%%%%%%%%%%%%%%%%%%%%%%%%%%%%%%%%%%%%%%%%%%%%%%%%%%%%%%%%%%%%%%
\label{II}

In this section we develop the OSEFT associated to hard or energetic photons. It fully corresponds to an effective field theory treatment of the so called eikonal or optical limit. The eikonal approximation is considered as a sort of semi-classical approach, valid when the wavelength of the photon is much shorter than any other length scale in the problem. The OSEFT allows us to study corrections to the pure classical term as a series of operators of increasing dimension over powers of the photon energy, which is the inverse of the photon wavelength. These will represent quantum corrections to the semi-classical picture, an explicit example of which will be presented in the manuscript.

Our starting point is the Lagrangian describing the propagation of free electromagnetic fields \footnote{It is possible to generalize this procedure for electromagnetic field propagation in a medium characterized by a refractive index, as long as the space of variation of the index is much larger than the photon wavelength. We will leave this case for future studies. }
\begin{equation}\label{L_start}
\mathcal{L}=-\dfrac{1}{4}F_{\mu\nu}F^{\mu\nu} \ .
\end{equation}
Here $F^{\mu\nu}(x)=\d^\mu A^\nu(x)-\d^\nu A^\mu(x)$ is the electromagnetic field strength tensor, and $A^\mu(x)$ is the vector gauge field potential. The momentum of an almost on-shell photon in the frame characterized by the time-like vector $u^\mu$ (satisfying $u^2=1$), can be decomposed into  on-shell and residual parts as follows
\begin{equation}\label{momentum_splitting}
q^\mu= p^\mu + k^\mu = Ev^\mu + k^\mu \ , 
\end{equation}
being $E= p \cdot u$  the energy of the photon in that frame, $v^\mu$ a light-like vector ($v^2=0$) and $k^\mu$ the so called residual momentum. The above decomposition assumes $k^\mu\ll Ev^\mu$, so that one can identify the photon energy $E$ as the hard scale. We define another light-like vector $\tilde{v}^\mu$ ($\tilde{v}^2 =0$) satisfying 
$v \cdot \tilde{v} =2$, such that 
\begin{equation}
u^\mu = \frac{ v^\mu +  \tilde{v}^\mu}{2} \ .
\end{equation}
Moreover, it will be useful to introduce an additional space-like vector 
\begin{equation}
n^\mu = \frac{ \tilde{v}^\mu-v^\mu}{2} \ ,
\end{equation}
which satisfies $n^2=-1$. Then, the momentum splitting of Eq.~\eqref{momentum_splitting} is performed at the Lagrangian level 
\begin{equation}\label{L_EFT}
	\La_{E, v}=-\dfrac{1}{4}\left( \d^\mu A^{\nu}_v -\d^\nu A^{\mu}_v \right)^2 \ ,
\end{equation}
and we factor out the hard momenta of the vector gauge field potential as
\begin{equation}\label{ansantz_photon_field}
A^{\mu}_v(x)= e^{-i E v\cdot x}\xi^\mu(x)+e^{i E v\cdot x}\xi^{\mu\dag}(x)\ .
\end{equation}
The above decomposition then assumes that the field $\xi^{\mu}(x)$ only has a dependence on the residual momenta. In terms of this field we can rewrite the Lagrangian of Eq.~\eqref{L_EFT} as 
\begin{equation}
\La_{E,v}= \dfrac{1}{2}\xi^{\mu\dag}\left( g_{\mu\nu}\square-\d_\mu\d_\nu-2iE g_{\mu\nu}(v\cdot \d)+E^2 v_\mu v_\nu+iE(v_\mu \d_\nu+\d_\mu v_\nu)\right) \xi^{\nu}  + \text{h.c.} \ ,
\end{equation}
where the oscillating terms $\sim e^{\pm 2iE v\cdot x}$ have been dropped, also, we use $\square=\d^\mu\d_\mu$ and h.c. stands for hermitian conjugate. However, we can as well construct an artificial covariant derivative by introducing 
\begin{equation}\label{derivative}
\D^\mu \equiv \d^\mu-iEv^\mu \ ,
\end{equation}
and rewrite the Lagrangian in the compact form
\begin{equation}\label{Lagrangian_tauv}
\La_{E,v}=  \dfrac{1}{2}\xi^{\mu\dag}  \left(  g_{\mu\nu}\D^2-\D_{\mu}\D_\nu\right) \xi^{\nu}  +\text{h.c} \ ,
\end{equation}
with $\D^2=\D^\mu \D_\mu=\square-2iE (v\cdot\d)$. The $\xi^\mu(x)$ field contains both physical and nonphysical components. At the classical level, this last can be eliminated for instance by imposing gauge fixing conditions. Below we will show that, under the above assumptions, the nonphysical components can be eliminated from the Lagrangian using effective field theory techniques, with no need of gauge-fixing. 

Let us start by introducing the transverse projector to $v^\mu$ and $\tilde{v}^\mu$
\begin{align}\label{Projector_perp_1}
& P_\perp ^{\mu\nu}=g^{\mu\nu}-\dfrac{1}{2}(v^\mu \tilde{v}^\nu+ \tilde{v}^\mu v^\nu) \ ,
\end{align}
which obeys
\begin{equation}
P_{\perp}^{\mu\nu}P_{\perp,\mu\rho}=P_{\perp,\rho}^{\nu} \ , \quad P_\perp^2=2 \ .
\end{equation}
Note that in the rest frame, where $u^\mu=(1,\boldsymbol{0})$, one has $v^\mu=(1,\v)$, $\tilde{v}^\mu=(1,-\v)$ and $n^\mu=(0,-\v)$, so that the transverse projector only has spatial components in that frame and is perpendicular to $\v$. Using the transverse projector, we can split the  $\xi^\mu(x)$ field into different components
\begin{equation}\label{tau_components}
\xi^\mu(x)=\xi_{\perp}^\mu(x)+v^\mu\phi (x)  +n^\mu\lambda (x)\ ,
\end{equation}
where each component is clearly identified
\begin{equation}
	\xi_{\perp}^\mu(x)=P_\perp ^{\mu\nu} \xi_{\nu}(x)  \ , \qquad \phi (x)=(u\cdot\tau )(x) \ , \qquad \lambda (x)=(v\cdot\tau )(x) \ .
\end{equation}
The $\xi_{\perp}^\mu(x)$ field  describes transverse degrees of freedom to $v^\mu$ and $\Tilde{v}^\mu$, while the remaining longitudinal and scalar degrees of freedom are both described by the $\lambda (x)$ and $\phi (x)$ fields. There is a certain freedom in the decomposition of the non-transverse part of Eq.\eqref{tau_components}. We have chosen the one that singles out the component that is still transverse to the frame vector $u^\mu$, that is $\lambda(x)$ (since $u\cdot n=0$). The suitability of this choice will be explained throughout the paper. In terms of these components, the Lagrangian can be written as
\begin{align} \label{Lagrangian_photons}  \notag 
		 \La_{E,v}=\dfrac{1}{2}\bigg\lbrace\xi_{\perp}^{\mu\dag} \left( g_{\mu\nu}\D^2-\d_\mu \d_\nu\right)  \xi_{\perp}^{\nu}
		-\phi^\dag  (v\cdot \d)^2  \phi 
        -\lambda^\dag (\D^2+(n\cdot \D)^2) \lambda  
		\\  \notag 
		-\xi_{\perp}^{\mu\dag} \d_\mu(v\cdot \d)  \phi 
        -\phi^\dag    (v\cdot \d) \d_\mu  \xi_{\perp}^{\mu}
        -\xi_{\perp}^{\mu\dag}\d_\mu(n\cdot \D)\lambda 
		-\lambda^\dag  (n\cdot \D)\d_\mu \xi_{\perp}^{\mu}
        \\
		  +\phi^\dag  (\D^2-(v\cdot\d)(n\cdot \D)) \lambda 
		+\lambda^\dag   (\D^2-(v\cdot\d)(n\cdot \D))\phi  \bigg\rbrace
		+\text{h.c}\ .
\end{align} 
Above, one can see that the operators accounting for the propagation of the various components do not have the same power counting in energy $E$, for instance $\D^2\sim E$ and $(n\cdot \D)^2\sim E^2$. Hence, we identify the $\lambda (x)$ field as the degree of freedom that can be integrated   out using its classical equation of motion
\begin{equation}\label{motion_lambda}
\lambda (x)=-\frac{ 1}{\D^2+(n\cdot\D)^2} \left[(n\cdot \D)\partial_\mu \xi^{\mu}_{\perp}(x)-(\D^2-(v\cdot\d)(n\cdot \D))\phi (x)\right] \ .
\end{equation}
The resulting Lagrangian may be written as
\begin{align}\label{L_integrated}  \notag 
&\La_{E,v}=
\dfrac{1}{2}\bigg\lbrace\xi_{\perp}^{\mu\dag} \, \left( g_{\mu\nu}\D^2 -\d_\mu\d_\nu\dfrac{\D^2}{\D^2+(n\cdot \D)^2}\right)  \xi_{\perp}^{\nu}
\\ 
& -\xi_{\perp}^{\mu\dag} \d_\mu\dfrac{\D^2(u\cdot\D)}{\D^2+(n\cdot\D)^2}\phi -\phi^\dag \dfrac{\D^2(u\cdot \D)}{\D^2+(n\cdot \D)^2}\d_\mu\xi_{\perp}^{\mu}
+\phi^\dag  \dfrac{\d_\perp^2 \D^2}{\D^2+(n\cdot\D)^2}  \phi \bigg\rbrace + \text{h.c} \ ,
\end{align}
where we defined $\d_\perp^{\mu} \equiv P_\perp^{\mu\nu}\d_\nu$. In the above Lagrangian and for the rest of the paper, we will use a compact operator notation that we elaborate in App.~\ref{A}. Subsequently, the above operators are expanded and organized in inverse powers of the hard scale $(1/E)^{n}$, yielding an infinite series of Lagrangians $\La_{E,v}^{(n)}$, each of them encompassing operators  that increase in dimension with the power of $n$. Specifically, using the expansions
\begin{subequations}
\begin{align}
\label{operator_expansion_initial}
& \dfrac{\D^2}{\D^2+(n\cdot \D)^2}= \dfrac{i(v\cdot\d)}{E}+\mathcal{O}\left(\dfrac{1}{E^2}\right) \ ,
\\
& \dfrac{\D^2(u\cdot\D)}{\D^2+(n\cdot \D)^2}= 2(v\cdot \d)+\dfrac{i}{E}\left(\d_\perp^2-(v\cdot\d)^2\right)+\mathcal{O}\left(\dfrac{1}{E^2}\right) \ ,
\end{align}
\end{subequations}
we can write down the the first three orders
\begin{subequations}\label{Lagrangian_expansion}
\begin{align}
& \La_{E,v}^{(-1)}=-E\, \xi_{\perp}^{\mu\dag} \, g_{\mu\nu}(iv\cdot \d) \xi_{\perp}^{\nu}+\text{h.c} \ ,
\\\label{Lagrangian_expansion_2}
& \La_{E,v}^{(0)}=\dfrac{1}{2}\xi_{\perp}^{\mu\dag} \, g_{\mu\nu}\square \xi_{\perp}^{\nu}
-\xi_{\perp}^{\mu\dag}\, \d_{\mu} \left( v\cdot \d\right) \phi 
-\phi^\dag \left( v\cdot \d\right)\d_{\mu}\xi_{\perp}^{\mu}+ \text{h.c} \ ,
\\ \notag
& \La_{E,v}^{(1)}=\dfrac{i}{E}\bigg(-\xi_{\perp}^{\mu\dag} \, \d_\mu(v\cdot \d) \d_\nu\xi_{\perp}^{\nu}-\dfrac{1}{2}\xi_{\perp}^{\mu\dag}\, \d_{\mu}  \left(\d_\perp^2-(v\cdot\d)^2\right)\phi -\dfrac{1}{2}\phi^\dag  \, \left(\d_\perp^2-(v\cdot\d)^2\right)\d_{\mu}\xi_{\perp}^{\mu} 
\\ \label{Lagrangian_expansion_3}
& 
+\phi^\dag  \, \d_\perp^2(v\cdot\d) \phi \bigg)
+\text{h.c}  \ .
\end{align}
\end{subequations}
Beyond the leading order, scalar and longitudinal degrees of freedom endure, indicated by the presence of terms with the $\phi (x)$ field. However, we can eliminate those terms through the application of local field redefinitions. Indeed, the following transformation
\begin{equation} \label{Local_field_redefinition_first_order}
	\xi^{\mu}_{\perp}(x) \ \longrightarrow \ \tau^{\mu}_{\perp}(x)=\xi_{\perp}^{\mu}(x)-\dfrac{i\d_\perp^\mu}{E}\phi (x)- \dfrac{(v\cdot\d+\Tilde{v}\cdot\d)\d_\perp^\mu}{2E ^2}\phi (x) \ ,
\end{equation}
completely eliminates all terms with the $\phi(x)$ field from Eqs.~(\ref{Lagrangian_expansion_2}-\ref{Lagrangian_expansion_3}). We have demonstrated here a specific case of a much broader scenario. In general, to eliminate the $\phi (x)$ field from the $n$-th order Lagrangian, one needs to apply the following field redefinition
\begin{equation}\label{LFR_all_orders}
	\xi_{\perp}^{\mu}(x)\ \longrightarrow \ \tau^{\mu}_{\perp}(x)=\xi_{\perp}^{\mu}(x) -\dfrac{\d_\perp^\mu}{u\cdot \D}\phi (x) \ ,
\end{equation}
with the operators expanded up to $(n+1)$-th order in inverse powers of $E$. In this way, the $\phi (x)$ field can be eliminated at all orders in the energy expansion and the final Lagrangian can be written in terms of the locally redefined field only as
\begin{equation}\label{Lagrangian_OSEFT}
\begin{gathered}
\La_{E,v}'=
\dfrac{1}{2}\tau _{\perp}^{\mu\dag} \, \left( g_{\mu\nu}\D^2 -\d_\mu \d_\nu\dfrac{\D^2}{\D^2+(n\cdot \D)^2}\right)  \tau _{\perp}^{\nu} +\text{h.c} \ .
\end{gathered}
\end{equation}
The final OSEFT Lagrangian only contains the vector degree of freedom $\tau_\perp^\mu(x)$, corresponding to transverse, almost on-shell photons. Note however, that we did not impose any particular gauge to derive it, in fact, we demonstrate in Sec.\ref{II.C} that it enjoys a gauge symmetry. Notably, the equivalence principle \cite{Haag:1958vt,Coleman:1969sm,Manohar:2018aog} guarantees that the on-shell quantities
after the local field redefinition of Eq.\eqref{LFR_all_orders} remain unaffected. 

Expanding now to the lowest orders we have
\begin{subequations}\label{Lagrangian_expansion_LFR}
\begin{align}
\label{Lagrangian_expansion_LFR_-1}
& \La_{E,v}^{ ,(-1)}=-E\tau_{\perp}^{\mu\dag} \, g_{\mu\nu}(iv\cdot \d) \tau _{\perp}^{\nu}+\text{h.c} \ ,
\\
\label{Lagrangian_expansion_LFR_0}
& \La_{E,v}^{ ,(0)}=\dfrac{1}{2}\tau_{\perp}^{\mu\dag} \, g_{\mu\nu}\square \tau _{\perp}^{\nu}+ \text{h.c} \ ,
\\ 
\label{Lagrangian_expansion_LFR_1}
& \La_{E,v}^{ ,(1)}=-\dfrac{1}{E}\tau_{\perp}^{\mu\dag} \, \d_\mu(i v\cdot \d) \d_\nu\tau _{\perp}^{\nu} + \text{h.c}\ .
\\ 
\label{Lagrangian_expansion_LFR_2}
& \La_{E,v}^{ ,(2)}=\dfrac{1}{2E^2}\tau_{\perp}^{\mu\dag} \, \d_\mu(\square-4(u\cdot\d)(v\cdot\d)) \d_\nu\tau _{\perp}^{\nu} + \text{h.c}\ .
\end{align}
\end{subequations}
The leading order Lagrangian describes the so called eikonal limit, while the higher order Lagrangians capture the corrections to this limit. Let us discuss how the above Lagrangian should be interpreted, as in an effective field theory, one has to solve the theory order by order. For example, naively, the equation of motion as derived from Eqs.(\ref{Lagrangian_expansion_LFR_-1}-\ref{Lagrangian_expansion_LFR_2}) reads\footnote{We multiplied by an overall factor of two for convenience.}
\begin{equation}
\left(-g_{\mu\nu}\left(2E\, iv\cdot \d -\square\right)-\d_\mu\left(\dfrac{2iv\cdot\d}{E}-\dfrac{\square-4(u\cdot\d)(v\cdot\d)}{E^2}\right)\d_\nu\right)\tau _{\perp}^{\nu}(x)=0 \ .
\end{equation}
However, one can always use the equation of motion at a given order in $1/E$ as a constraint for the operators appearing at the next order. For instance, the equation of motion at leading order $i(v\cdot\d)\tau_{\perp}^{\mu}(x)=0$ can be used to simplify the operator $\square \tau_{\perp}^{\mu}(x) =  \d_\perp^2\tau_{\perp}^{\mu}(x) + {\cal O} (1/E)$ appearing at the next order. Performing this process at a fixed order in $1/E$, one can always show without much difficulty that all structures proportional to $\d^\mu \d^\nu$ vanish, so that the equation of motion only contains the tensor structure $g^{\mu\nu}$ and can be formally written as
\begin{equation}\label{motion_tau_perp'}
-\left(2E\, iv\cdot \d -\square\right) \tau_{\perp}^{\mu}(x)=0 \ ,
\end{equation}
or as $\D^2\tau_{\perp}^{\mu}(x)=0$, by employing the operator in Eq.\eqref{derivative}. 

%%%%%%%%%%%%%%%%%%%%%%%%%%%%%%%%%%%%%%%%%%%%%%%%%%%%%%%%%%%%
\subsection{Vacuum propagator and dispersion relation}
%%%%%%%%%%%%%%%%%%%%%%%%%%%%%%%%%%%%%%%%%%%%%%%%%%%%%%%%%%%%
\label{II.A}

The OSEFT Green function in vacuum, that we define as $\mathcal{G}_\perp^{\mu\nu}(x,y)=\braket{0\vert\mathcal{T}\tau_{\perp}^{\mu\dag}(x)\tau_{\perp}^{\nu}(y)\vert 0}$, where $\mathcal{T}$ denotes time-ordering, obeys
\begin{equation}\label{motion_Green}
\D^2_x \, \mathcal{G}_\perp^{\mu\nu}(x,y)=P_\perp^{\mu\nu}\delta(x-y) \ ,
\end{equation}
as can be deduced from Eq.\eqref{motion_tau_perp'}. The above equation can be easily inverted in (residual) momentum space, yielding to
\begin{equation}
\mathcal{G}_{\perp}^{\mu\nu}(k)= \dfrac{-P_\perp^{\mu\nu}}{2E(v\cdot k)+k^2+i0^+} \ .
\end{equation}
The dispersion equation is $2E(v\cdot k)+k^2=0$, which has to be solved order by order in the energy expansion (see for instance Refs.\cite{Manuel:2014dza,Manuel:2016wqs,Manuel:2016cit}). Doing so up to order $1/E$, one gets
\begin{equation}\label{eq_dis_NNLO}
2E(v\cdot k)+ k_\perp^2-\dfrac{(\Tilde{v}\cdot k) k_\perp^2}{2E}=0 \ ,
\end{equation}
so that in the rest frame, using the notation $k^\mu=(k_0,\k)$ for the residual momentum, the denominator of the propagator can be written as $2E(k_0-\mathrm{f}(\k))+i0^+$ where $\mathrm{f}(\k)$ is the dispersion relation of the OSEFT, given by
\begin{align}
\mathrm{f}(\k)= \v\cdot \k+\dfrac{\k_\perp^2}{2E}-\dfrac{(\v\cdot \k) \k_\perp^2}{2E^2}+ \mathcal{O}\left(\dfrac{1}{E^3}\right) \ ,
\end{align}
where we used the fact that in the rest frame one has $k_\perp^2=-\k_\perp^2$ and $\Tilde{v}\cdot k\approx 2(\v\cdot \k)$.

%%%%%%%%%%%%%%%%%%%%%%%%%%%%%%%%%%%%%%%%%%%%
\subsection{Effective gauge field for almost on-shell photons}
%%%%%%%%%%%%%%%%%%%%%%%%%%%%%%%%%%%%%%%%%%%%
\label{II.B}

It is interesting to relate our starting effective field $\xi^{\mu}(x)$ with the locally redefined field of the OSEFT Lagrangian $\tau_\perp^{\mu}(x)$. In order to do so, we first insert the equation of motion of the $\lambda (x)$ field into Eq.\eqref{tau_components}, in addition, by writing $\d_\perp^\mu=\d^\mu-v^\mu (u\cdot\d)-n^\mu(v\cdot\d)$ in the field redefinition of Eq.\eqref{LFR_all_orders} and using the operator identity of Eq.\eqref{operator_identity}, we find that an effective gauge field for almost on-shell photons is
\begin{align}\label{relation_tau}
\xi^\mu(x)=\left(g^\mu_\nu-n^\mu\d_\nu\frac{ n\cdot\D}{\D^2+( n\cdot\D)^2}\right)\tau _{\perp}^{\nu}(x) +\dfrac{\D^\mu}{u\cdot \D}\phi (x) \ .
\end{align}
Actually,  it is relatively easy to derive again the OSEFT Lagrangian in very few steps, by directly plugging the above relation into Eq.\eqref{Lagrangian_tauv} and noting that all terms with the $\phi (x)$ field vanish due to $\left( g_{\mu\nu}\D^2-\D_{\mu}\D_\nu\right) \D^\mu=0 $. 

The above expression is quite useful in computations, for instance, in Sec.\ref{III} we will use it to construct the photon Wigner function in a semi-classical approximation,
furthermore, it will help to understand the physical picture underlying the OSEFT. This last question will be elaborated throughout the manuscript, in particular in the next section, where we discuss how the gauge symmetry is realized in the effective field theory.

%%%%%%%%%%%%%%%%%%%%%%%%%%%%%%%%%%%%%%%%%%%%%%%%%%%%%%%%%%%%
\subsection{OSEFT gauge transformations}
%%%%%%%%%%%%%%%%%%%%%%%%%%%%%%%%%%%%%%%%%%%%%%%%%%%%%%%%%%%%
\label{II.C}

In effective field theories, where there is a well defined hierarchy of scales, it is common to separate the gauge transformations for each sector of the theory according to their scale, respecting such a separation. One particular example of this fact occurs  in SCET, where one talks on a gauge symmetry associated to the hard (or collinear) gluon fields, and another one associated to the  soft gluon fields \cite{Becher:2014oda}. An additional multipole expansion of the different fields might be needed to respect the energy separation \cite{Beneke:2002ni}. An analogous situation is expected in the OSEFT for photons. In this work, we only consider the hard sector of the theory, ignoring the soft gauge fields,  as they do not interact. Further, we are not considering the interaction with matter particles neither. All of them could be incorporated, following the same SCET techniques.  Here we will discuss the gauge symmetry associated to hard photons. The Lagrangian of Eq.\eqref{L_EFT} enjoys the gauge symmetry
\begin{equation}\label{gauge_trans}
A^\mu_v (x)\longrightarrow A^\mu_v (x)+\d^\mu \theta_v (x) \ ,
\end{equation}
for an arbitrary function $\theta_v (x)$ that respects the energy scaling of the gauge field. Hence, the OSEFT Lagrangian of Eq.\eqref{Lagrangian_OSEFT} should also posses a gauge symmetry. By writing $\theta_v (x)=e^{-iE v\cdot x}\eta (x)+\text{h.c}$, we can deduce how the different components of the photon field transform under a gauge transformation. Explicitly, one then finds that the OSEFT vector gauge field transforms as
\begin{equation}
\xi^{\mu}(x)\longrightarrow \xi^{\mu}(x) +\D^\mu\eta (x)\ ,
\end{equation}
and thus
\begin{subequations}
\begin{align}
&\xi_{\perp}^{\mu}(x)\longrightarrow\xi_{\perp}^{\mu}(x)+\d^\mu_\perp \eta (x) \ , 
\\ 
& \phi (x)\longrightarrow \phi (x)+(u\cdot \D)\eta (x)  \ , 
\\
& \lambda (x)\longrightarrow \lambda (x)+(v\cdot\d)\eta (x)  \ .    
\end{align}    
\end{subequations}
Then, we can show that the final OSEFT Lagrangian is gauge invariant. To prove it, we note that the locally redefined field is itself invariant under the above set of gauge transformations. Indeed,
\begin{equation}
\tau_{\perp}^{\mu}(x)\longrightarrow\xi_{\perp}^{\mu}(x) +\d^\mu_\perp \eta (x)-\dfrac{\d_\perp^\mu}{u\cdot \D}\left[\phi (x)+(u\cdot \D)\eta (x)\right]=\tau_{\perp}^{\mu}(x)\ .
\end{equation}
%
%It is also interesting to see how the effective gauge field for almost on-shell photons transforms under a gauge transformation. Taking into account these last remarks, from Eq.\eqref{relation_tau} we easily find
%
%\begin{equation}
%\xi^{\mu}(x)\longrightarrow \xi^{\mu}(x) +\D^\mu\eta (x)\ ,
%\end{equation}
%
%so that it is not a gauge invariant quantity, as expected. Hence, in the OSEFT, the (gauge) field $\xi^{\mu}(x)$ is decomposed into the fields $\tau_{\perp}^{\mu}(x)$ and $\phi(x)$, the former is gauge invariant and the later is not. Another peculiarity of the theory is that the aforementioned components do not couple to each other.

Let us conclude this section with the following observation. Taking the four divergence in Eq.\eqref{ansantz_photon_field} and resorting to Eq.\eqref{relation_tau} one can show that
\begin{equation}\label{Lorentz_condition}
\d_\mu A^\mu_v=\dfrac{\D^2}{(u\cdot\D)}\phi(x)e^{-i E v\cdot x}+\text{h.c} \ .
\end{equation}
Thus, if we had not carried out the local field redefinition and rather we would have fixed the gauge as $\phi(x)=0$ in the initial Lagrangian, that would imply that the gauge field obeys $u_\mu A_v^\mu=0$ and $\d_\mu A^\mu_v=0$. In conclusion, the resulting framework in the OSEFT if we impose that $\phi(x)=0$ is equivalent to the Coulomb gauge. However, let us stress that in this work we perform a local field redefinition to eliminate the non-physical $\phi(x)$ component of the gauge field from the Lagrangian, which allows us to work without choosing any particular gauge.

%%%%%%%%%%%%%%%%%%%%%%%%%%%%%%%%%%%%%%%%%%%%
\subsection{Polarization vectors}
%%%%%%%%%%%%%%%%%%%%%%%%%%%%%%%%%%%%%%%%%%%%
\label{II.D}

The field $\tau_{\perp}^{\mu}(x)$ is transverse to $v^\mu$ and $\Tilde{v}^\mu$. Thus, in the rest frame, it only has spatial components and is transverse to $\v$. It is suggesting then to introduce polarization vectors in the effective field theory.  
\\
Let us define a linear polarization basis, as $\lbrace e_{i}^\mu\rbrace$ with $i=\lbrace 1, 2\rbrace$, satisfying $(e_i^\mu)^*=e_i^\mu$, also 
\begin{equation}
v\cdot e_i=\Tilde{v}\cdot e_i=0 \ , \quad e_i\cdot e_j=g_{ij} =-\delta_{ij} \ ,
\end{equation}
so that they are unitary and space-like. 
 We shall, however, work with circular polarization vectors, that we introduce as $\lbrace e_h^\mu\rbrace$ with $h=\lbrace+,-\rbrace $, obeying the properties
\begin{equation}
v\cdot e_h=\Tilde{v}\cdot e_h=0 \ , \quad e_h^*\cdot e_{h'}=-\delta_{hh'} \ .
\end{equation}
Their relation with the linear polarization vectors is
\begin{equation}
e^\mu_h=\dfrac{1}{\sqrt{2}}(e_1^\mu +ih\,e_2^\mu) \ ,
\end{equation}
so that $(e_h^\mu)^*=e_{-h}^\mu$. Note that both the linear and circular polarization vectors are also transverse to $u^\mu$ and $n^\mu$. We can relate the polarization vectors with the transverse projector and spin tensor (introduced below) as
\begin{subequations}
\begin{align}
& P_\perp^{\mu\nu}=g^{\mu\nu}-\dfrac{1}{2}(v^\mu \tilde{v}^\nu+ \tilde{v}^\mu v^\nu)= -e^{*\mu}_+ e^{\nu}_+ -e^{*\mu}_{-} e^{\nu}_{-} \ ,    
\\ \label{spin_tensor_OSEFT}
& S_\perp^{\mu\nu}=i\eps^{\mu\nu\alpha\beta}v_\alpha u_\beta=e^{*\mu}_+ e^{\nu}_+ - e^{*\mu}_- e^{\nu}_- \ .
\end{align}    
\end{subequations}
The spin tensor obeys $S_\perp^2=-2$, but it is not a good projector, since $S_{\perp}^{\mu\nu}S_{\perp,\mu\rho}=-P_{\perp,\rho}^{\nu}$.  Thus, it is useful to introduce the right ($h=+$) and left ($h= -$) projectors
\begin{align}\label{P_h}
 P_{h}^{\mu\nu}= \dfrac{1}{2}(P_\perp^{\mu\nu}-h S_\perp^{\mu\nu} )= -e^{*\mu}_h e^{\nu}_h \ , 
\end{align}
satisfying the properties
\begin{align}
P_{\pm}^2=0 \ , \quad  P_{\pm}^{\mu\nu}P_{\mp,\mu\nu}=1 \ , \quad P_{\pm}^{\mu\nu}P_{\pm,\mu\rho}=0 \ , \quad P_{\pm}^{\mu\nu}P_{\mp,\mu\rho}=-P_{\mp,\rho}^{\nu} \ .
\end{align}
Now, the field $\tau_{\perp}^{\mu}(x)$ can be decomposed in the circular polarization basis as
\begin{equation}
\tau_{\perp}^{\mu}(x)=\sum_{h=\pm}e_h^\mu \tau_h  (x) \ .
\end{equation}
Hence, the components $\tau_h  (x)=-(e_h^*\cdot \tau_\perp  )(x)$ correspond to right $(h=+)$ and left $(h=-)$ handed circularly polarized photons respectively.

%%%%%%%%%%%%%%%%%%%%%%%%%%%%%%%%%%%%%%%%%%%%%%%%%%%%%%%%%%%%
\subsection{Reparametrization invariance}
%%%%%%%%%%%%%%%%%%%%%%%%%%%%%%%%%%%%%%%%%%%%%%%%%%%%%%%%%%%%
\label{II.E}

In general, an effective field theory should remain invariant under infinitesimal transformations of the parameters used to describe the theory. The invariance under these types of transformations is called reparametrization invariance (RI) and it was first discussed in the context of heavy quark effective field theory (HQEFT) \cite{Luke:1992cs}, and later on generalized for massless degrees of freedom in the context of soft collinear effective field theory (SCFT) \cite{Manohar:2002fd}. 

RI is the symmetry associated with the ambiguity of the decomposition of the full momentum $q^\mu$  in Eq.~(\ref{momentum_splitting}). A small shift in the velocity $v^\mu$ could be reabsorbed in the definition of the residual momentum $k^\mu$, while the physics should remain unchanged. On the other hand, the explicit choices of the vectors $v^\mu$  and ${\tilde v}^\mu$ seem to imply an apparent breaking of the Lorentz symmetry. Checking the RI of the theory ultimately confirms that Lorentz symmetry is respected in the effective field theory. 

RI has been extensively studied also in the OSEFT for fermions in \cite{Carignano:2018gqt,Manuel:2021oah}. 
We will discuss how to generalize these concepts for the photon fields here.
The main idea is that the field $A_v^\mu(x)$ should not change under infinitesimal transformations of the parameters used to construct the effective field theory. Then, under a RI transformation one should have
\begin{equation}
A_v^\mu(x) \overset{\Lambda}{\longrightarrow } A_{v'}^{ \mu}(x)=A_v^\mu(x)\ ,
\end{equation}
where we introduced the label $\Lambda=\lbrace \text{I, II, III}\rbrace $ for the three types of RI transformations  \cite{Manohar:2002fd}. Thus, the OSEFT Lagrangian should enjoy the following symmetries
\begin{equation}\label{RI_transformations}
	\text{(I)}:
	\begin{cases}
		v^\mu\rightarrow v^\mu +\Delta_\perp^\mu \\
		\Tilde{v}^\mu \rightarrow \Tilde{v}^\mu
	\end{cases}
    \ , \quad 
	\text{(II)}:
	\begin{cases}
		v^\mu\rightarrow v^\mu  \\
		\Tilde{v}^\mu \rightarrow \Tilde{v}^\mu +\tilde{\Delta}_\perp^\mu
	\end{cases}
    \ , \quad 
	\text{(III)}:
	\begin{cases}
		v^\mu\rightarrow (1+\alpha)v^\mu  \\
		\Tilde{v}^\mu \rightarrow (1-\alpha)\Tilde{v}^\mu
	\end{cases}
 \ ,
\end{equation}
where $\lbrace \Delta_\perp^\mu,\tilde{\Delta}_\perp^\mu,\alpha\rbrace$ are five infinitesimal parameters, satisfying 
\begin{equation}
\Delta_\perp\cdot v=\Delta_\perp\cdot \Tilde{v}=\Tilde{\Delta}_\perp\cdot v=\Tilde{\Delta}_\perp\cdot \Tilde{v}=0 \ .
\end{equation}
In Tab.(\hyperref[RI_table_start]{I}-\hyperref[RI_table_operators]{II}) we show the transformation rules for the projected fields and other relevant quantities of the effective field theory respectively. Polarization vectors also change under RI transformations, as can be seen in the last row of Tab.(\hyperref[RI_table_operators]{II}), so that they preserve their transversality to $v^\mu$ and $\Tilde{v}^\mu$.
\begin{table}[h] \label{RI_table_start}
	\begin{tabular}{|M{1cm}||M{5.8cm}|M{4.5cm}|M{3cm}|}
		\hline  
		\ & Type (I) & Type (II) & Type (III)\\ [0.8ex]
		\hline
		   $\xi_{\perp}^{\mu}$ & 
		  $\xi_{\perp}^{\mu}-\Delta_\perp^\mu \phi +\frac{1}{2} \Delta_\perp^\mu \lambda  -\frac{1}{2}v^\mu(\Delta_\perp\cdot \tau)  $
		& 
		$\xi_{\perp}^{\mu}-\frac{1}{2}v^\mu(\Tilde{\Delta}_\perp\cdot \tau)
	      -\frac{1}{2}\tilde{\Delta}_\perp^\mu\lambda $
		& 
		$\xi_{\perp}^{\mu}$ 
		\\ [1.2ex]
		\hline 
		   $\phi$ &  $\phi+\frac{1}{2}(\Delta_\perp\cdot\tau)$ &  $\phi+\frac{1}{2}(\Tilde{\Delta}_\perp\cdot\tau)$   & $(1-\alpha)\phi+\alpha \lambda$ \\ [1.2ex]
		\hline 
		  $\lambda$ &  $\lambda+\Delta_\perp\cdot\tau$  & $\lambda$ & $(1+\alpha)\lambda$ 
        \\[1.2ex]
		\hline
	\end{tabular}
	\caption{Transformation rules for the projected fields $\tau _{\perp}^\mu(x)$, $\lambda (x)$ and $\phi (x)$. We dropped the space time arguments to enlighten the notation.}
\end{table}

After integrating out, the transformations in Tab.(\hyperref[RI_table_start]{I}) with the presence of the $\lambda(x)$ field have to be modified. We derive and discuss the general expressions for those transformations, i.e valid at any order in $1/E$, in App.~\ref{B}. Remarkably, it can also be shown that the OSEFT Lagrangian, obtained after the local field redefinition to eliminate the $\phi(x)$ field, is RI invariant. In order to prove it, one just needs to know how the field $\tau_{\perp}^{\mu}(x)$, introduced in Eq.\eqref{LFR_all_orders}, changes under each type of transformation. We also give the general form of these last transformations in App.~\ref{B}. 

Hence, using Eqs.(\ref{RI_type_I_final}-\ref{RI_type_III_final}), it is possible to check that
%we see that 
%
\begin{equation}
\delta_{(\text{I})}\mathcal{L}  _{E,v}=\delta_{(\text{II})}\mathcal{L}  _{E,v}=\delta_{(\text{III})}\mathcal{L}  _{E,v}=0 \ ,
\end{equation}
or in other words, the OSEFT Lagrangian is RI invariant. It is worth mentioning that all non-transverse structures in these last transformations are not necessary to proof the RI invariance of the Lagrangian, as those pieces always vanish when contracted with transverse tensors. However, they would be necessary to derive the transformation rules for other quantities, such as currents. The general transformations for $\tau_{\perp}^{\mu}(x)$ are quite simple when expanded in powers of $1/E$. Precisely, keeping terms up to first order in the energy expansion we find 
\begin{subequations}
\begin{align}\label{RI_I_tau_LFR_1/E}
& \tau_{\perp}^{\mu} \overset{\text{(I)}}{\longrightarrow } \tau_{\perp}^{\mu}-\dfrac{i\d_\perp^\mu}{2E}(\Delta_\perp\cdot\tau_\perp  )-\dfrac{i\Delta_\perp^\mu}{2E}(\d\cdot\tau_\perp  )-\dfrac{\Tilde{v}^\mu}{2}(\Delta_\perp\cdot \tau_\perp) \ ,
\\ \label{RI_II_tau_LFR_1/E}
& \tau_{\perp}^{\mu} \overset{\text{(II)}}{\longrightarrow } \tau_{\perp}^{\mu}-\dfrac{i\d_\perp^\mu}{2E}(\Tilde{\Delta}_\perp\cdot\tau_\perp  )+\dfrac{i\Tilde{\Delta}_\perp^\mu}{2E}(\d\cdot\tau_\perp  )-\frac{v^\mu}{2}(\Tilde{\Delta}_\perp\cdot \tau_\perp  )\ ,
\\ \label{RI_III_tau_LFR_1/E}
& \tau_{\perp}^{\mu} \overset{\text{(III)}\,}{\longrightarrow } \tau_{\perp}^{\mu}\ .
\end{align}   
\end{subequations}
Employing the expressions above, we derive in Sec.(\ref{III}) the transformation rules for the distribution function for photons at $1/E$ accuracy, in particular using type (II) transformations one can dervie the so called side jump effect.
\begin{table}[h] \label{RI_table_operators}
	\begin{tabular}{|M{1cm}||M{5.3cm}|M{5.3cm}|M{3cm}|}
		\hline  
		\ & Type (I) & Type (II) & Type (III)\\ [0.8ex]
		\hline  
		$E$ & $E$ & $E$ & $(1-\alpha)E$  \\ [1.2ex]
		\hline
		$\d^\mu$ & $\d^\mu+iE \Delta_\perp^\mu$  & $\d^\mu$ &  $\d^\mu$
		\\ 
		\hline
		$\d_\perp^\mu$ & 
		$\d_\perp^\mu -\frac{1}{2}(\Delta_\perp\cdot \d)\Tilde{v}^\mu
	    -\frac{1}{2}(\Tilde{v}\cdot \D)\Delta_\perp^\mu $ 
		&
		$
		\d_\perp^\mu
			-\frac{1}{2}(\Tilde{\Delta}_\perp\cdot \d)v^\mu
	    -\dfrac{1}{2}(v\cdot \d)\tilde{\Delta}_\perp^\mu $
		&
		$\d_\perp^\mu$
		\\ [1.2ex]
		\hline
		$\D^\mu$ & $\D^\mu$  & $\D^\mu$ & $\D^\mu$
		\\ [1.2ex]
		\hline
		$u\cdot\D$  &  $u\cdot\D+\frac{1}{2}(\Delta_\perp\cdot \d)$  & $u\cdot\D+\frac{1}{2}(\Tilde{\Delta}_\perp\cdot \d)$ & $u\cdot\D-\alpha(n\cdot\D)$\\ [1.2ex]
		\hline
		$n\cdot\D$  &  $n\cdot\D-\frac{1}{2}(\Delta_\perp\cdot \d)$  & $n\cdot\D+\frac{1}{2}(\Tilde{\Delta}_\perp\cdot \d)$ & $n\cdot\D-\alpha(u\cdot\D)$ \\ [1.2ex]
		\hline
		$\d_\perp^2$ & $\d_\perp^2-(\Delta_\perp \cdot \d)(\Tilde{v}\cdot \D)$  & $\d_\perp^2-(\Tilde{\Delta}_\perp \cdot \d)(v\cdot \d)$ & $\d_\perp^2$
		\\ [1.2ex]
		\hline
         $e_h$ & 
		  $e_h-\frac{1}{2}(e_h\cdot\Delta_\perp )\Tilde{v}^\mu $
		& 
		$e_h-\frac{1}{2}(e_h\cdot\Tilde{\Delta}_\perp )v^\mu $
		& 
		$e_h $ 
		\\ [1.2ex]
		\hline
	\end{tabular}
	\caption{Transformation rules for the operators of the effective field theory and the OSEFT polarization vectors. The transformation rules for the negative energy sector of the theory can be esasily recovered after replacing $E\rightarrow -E$ and $h\rightarrow -h$.}
\end{table}
%

%%%%%%%%%%%%%%%%%%%%%%%%%%%%%%%%%%%%%%%%%%%%
\subsection{Going backward to the full theory variables}
%%%%%%%%%%%%%%%%%%%%%%%%%%%%%%%%%%%%%%%%%%%%
\label{II.F}

When constructing the OSEFT, we decomposed the photon momentum $q^\mu$ as in  Eq.\eqref{momentum_splitting}, introducing the effective field theory variables $E,v^\mu$ and $k^\mu$. Any quantity computed from the OSEFT depends on these variables. However, it is desirable to be able to re-express the results in terms of the full theory momentum $q^\mu$. 

This is a well established process which has been extensively discussed in the literature of the OSEFT, see e.g Refs.\cite{Manuel:2016wqs,Manuel:2016cit,Carignano:2018gqt,Carignano:2019zsh,Manuel:2021oah}, so we just recall here the relevant expressions needed for this work. First, we will need the expression of the on-shell velocity $v^\mu_q$ in the effective field theory, which reads
\begin{equation}\label{OS_velocity}
v^\mu_q =  v^\mu+\dfrac{k_\perp^\mu}{E}-\dfrac{(\Tilde{v}\cdot k) k_\perp^\mu}{2E^2}-\dfrac{k_\perp^2 n^\mu}{2E^2} + {\cal O} \left(\frac{1}{E^3}\right) \ .
\end{equation}
It follows that $\Tilde{v}^\mu_q= 2u^\mu-v_q^\mu$, while the frame vector $u^\mu$ does not change when moving back to the full theory. Also, it is useful to define a space-like vector in the full theory, such that
\begin{align}
n^\mu_q =\dfrac{\Tilde{v}^\mu_q-v^{\mu}_q}{2} = n^\mu-\dfrac{k_\perp^\mu}{E}+\dfrac{(\Tilde{v}\cdot k)k_\perp^\mu }{2E^2}+\dfrac{k_\perp^2n^\mu}{2E^2} + {\cal O} \left(\frac{1}{E^3}\right) \ .
\end{align}
Let us introduce the polarization vectors of the full theory in the circular basis as $\lbrace e_{q,h}^{\mu}\rbrace $ with the suffix $(q)$ indicating that they are functions of momentum and $h=\lbrace +,-\rbrace$. We can relate them with the OSEFT polarization vectors $\lbrace e_h^\mu\rbrace $ using the following trick. By requiring that at each order in $1/E$
\begin{equation}
u\cdot e_{q,h} =v_q\cdot e_{q,h}=\Tilde{v}_q\cdot e_{q,h}=0\ , \quad e_{q,h}\cdot e_{q,h'}=-\delta_{hh'} \ ,
\end{equation}
holds, it is not difficult to realize that
\begin{equation}\label{P_h_pol_vectors}
e_{q,h}^\mu = e_h^\mu-\dfrac{e_h\cdot k_\perp}{E}n^\mu+\dfrac{(e_h\cdot k_\perp)}{2E^2}\left(k_\perp^\mu+(\Tilde{v}\cdot k)n^\mu \right) + {\cal O} \left(\frac{1}{E^3}\right) \ .
\end{equation}
Then, we can introduce the transverse projector and spin tensor of the full theory as
\begin{subequations}
\begin{align}\label{P_perp_full}
& P_{\perp,q}^{\mu\nu}= g^{\mu\nu}-\dfrac{1}{2}(v^\mu_q \Tilde{v}^\nu_q+\Tilde{v}^\mu_q v^\nu_q)=-e_{q,+}^{*\mu}e_{q,+}^{\nu}-e_{q,-}^{*\mu}e_{q,-}^{\nu} \ ,
\\ \label{S_perp_full}
& S_{\perp,q}^{\mu\nu}=\dfrac{ i\eps^{\mu\nu\alpha\beta} q_{\alpha} u_\beta}{u\cdot q}=e_{q,+}^{*\mu}e_{q,+}^{\nu}-e_{q,-}^{*\mu}e_{q,-}^{\nu} \ ,
\end{align}    
\end{subequations}
respectively. In the definitions above, $q^\mu=(E_q,\q)$ is the on-shell momentum (with $E_q=u\cdot q$), also $v^\mu_q=(1,\q/E_q)$ and $\tilde{v}^\mu_q=(1,-\q/E_q)$. They can also be expressed in terms of the OSEFT variables, either by using Eq.\eqref{OS_velocity} or Eq.\eqref{P_h_pol_vectors}. For instance, we can write
\begin{subequations}
\begin{align}\label{P_perp_full_EFT} 
& P_{\perp,q}^{\mu\nu}= P_\perp^{\mu\nu}-\dfrac{1}{E}n^{(\mu} k_\perp^{\nu)}+\dfrac{(\tilde{v}\cdot k)}{2E^2} n^{(\mu} k_\perp^{\nu)} +\dfrac{1}{E^2}(k_\perp^\mu k_\perp^\nu+k_\perp^2 n^\mu n^\nu) + {\cal O} \left(\frac{1}{E^3}\right)\ ,
\\ \label{S_perp_full_EFT}
& S_{\perp,q}^{\mu\nu}= S_\perp^{\mu\nu}+\dfrac{1}{E}n^{[\mu} S_{\perp}^{\nu]\alpha}  k_{\alpha}-\dfrac{1}{2E^2}k_\perp^{[\mu}S_\perp^{\nu]\alpha}k_\alpha-\dfrac{(\tilde{v}\cdot k)}{2E^2}n^{[\mu}S_\perp^{\nu]\alpha} k_\alpha  + {\cal O} \left(\frac{1}{E^3}\right)\ .
\end{align}    
\end{subequations}
% 

%%%%%%%%%%%%%%%%%%%%%%%%%%%%%%%%%%%%%%%%%%%%%%%%%%%%%%%%%%%%
\section{Quantum kinetic theory for photons from the OSEFT}
%%%%%%%%%%%%%%%%%%%%%%%%%%%%%%%%%%%%%%%%%%%%%%%%%%%%%%%%%%%%
\label{III}

In the previous section we derived the OSEFT associated to an energetic photon field. Here we will consider a many body system, such as a plasma, characterized by a temperature $T$. Then, one should consider that there are many  electromagnetic fields with energy scales of the order or much larger than $T$, which then admit an OSEFT description.
The electromagnetic fields with typical energy scales lower than $T$ can be then treated as classical gauge fields. This can be justified by the fact that the Bose-Einstein distribution function is well approximated by a classical field distribution function at low energies.

In this section we derive photon quantum kinetic equations 
from the effective field theory developed in Sec.~\ref{II}, that is, we assume that we are describing the high energetic photons in the system. To this aim, apart from the OSEFT for photons we employ the Schwinger-Keldysh formalism of thermal field theory \cite{Kapusta:2006pm,Bellac:2011kqa,Laine:2016hma}, see also Refs.\cite{Ghiglieri:2020dpq,Mustafa:2022got} for recent reviews on the subject. 

Let us emphasize that in the OSEFT, we have performed a local field redefinition to eliminate the non-physical component of the photon field from the Lagrangian (c.f Eq.\eqref{LFR_all_orders}). Notably, as the equivalence theorem was extended to finite density and then to general thermodynamic observables in Ref.\cite{Furnstahl:2000we}, doing a similar reasoning as in Sec.\ref{II} allows us to work with the locally redefined field with no need to worry about affecting any on-shell quantity. 

The main object in the Schwinger-Keldysh formalism is the Green function, which is expressed as a matrix in the complex time path contour. Similarly, in the OSEFT, one can define for positive energy photons
\begin{equation}\label{RT_Green_pos}
\boldsymbol{\cal G}^{\mu\nu}_{\perp}(x,y)= 
\begin{pmatrix}
{\cal G}^{c,\mu\nu}_{\perp}(x,y) & {\cal G}^{<,\mu\nu}_{\perp}(x,y)\\
{\cal G}^{>,\mu\nu}_{\perp}(x,y)& {\cal G}^{a,\mu\nu}_{\perp}(x,y)
\end{pmatrix}
= 
\begin{pmatrix}
\braket{\mathcal{T}_C\,\tau_{\perp}^{\mu}(x)\tau_{\perp}^{\nu\dag}(y)} &
\braket{\tau_{\perp}^{\nu\dag }(y) \tau_{\perp}^{\mu}(x)} \\
\braket{\tau_{\perp}^{\mu}(x)\tau_{\perp}^{\nu\dag}(y)} &
\braket{\widetilde{\mathcal{T}}_C\, \tau_{\perp}^{\mu}(x) \tau_{\perp}^{\nu\dag }(y) } 
\end{pmatrix}
\ ,
\end{equation}
where $\mathcal{T}_C$ and $\widetilde{\mathcal{T}}_C$ denote time and anti-time ordering along the complex time contour respectively, while $\braket{\ldots}$ denotes thermal average over an ensemble of states. Analogous definitions can be done for the negative energy sector of the theory. Indeed, we can also define
\begin{equation}\label{RT_Green_neg}
\boldsymbol{\widetilde{\cal G}}^{\mu\nu}_{\perp}(x,y)= 
\begin{pmatrix}
\widetilde{\cal G}^{c,\mu\nu}_{\perp}(x,y) & \widetilde{\cal G}^{<,\mu\nu}_{\perp}(x,y)\\
\widetilde{\cal G}^{>,\mu\nu}_{\perp}(x,y)& \widetilde{\cal G}^{a,\mu\nu}_{\perp}(x,y)
\end{pmatrix}
= 
\begin{pmatrix}
\braket{\mathcal{T}_C\,\tau_{\perp}^{\mu\dag}(x)\tau_{\perp}^{\nu}(y)} &
\braket{\tau_{\perp}^{\nu}(y) \tau_{\perp}^{\mu\dag}(x)} \\
\braket{\tau_{\perp}^{\mu\dag}(x)\tau_{\perp}^{\nu}(y)} &
\braket{\widetilde{\mathcal{T}}_C\, \tau_{\perp}^{\mu\dag}(x) \tau_{\perp}^{\nu}(y) } 
\end{pmatrix}
\ .
\end{equation}
Our interest is in the lesser (or greater) components of the Green functions, as these are related with the photon phase-space distribution function after a Wigner transform. From their definition, it follows that the lesser and greater components of each sector of the theory are related by
\begin{align}\label{relation_GF_pos_neg}
{\cal G}^{<,\mu\nu}_{\perp}(x,y)=\widetilde{\cal G}^{>,\nu\mu}_{\perp}(y,x)  \ ,
\end{align}
and satisfy the hermiticity property
\begin{align}\label{hermicity_properties_GF}
({\cal G}^{<,\mu\nu}_{\perp}(x,y))^*=\widetilde{\cal G}^{>,\mu\nu}_{\perp}(x,y)={\cal G}^{<,\nu\mu}_{\perp}(y,x)  \ .
\end{align}
Let us focus on the positive energy sector of the theory. We define the OSEFT Wigner function for positive energy photons as
\begin{align} \label{OSEFT_WF_pos}
{\cal G}^{<,\mu\nu}_{\perp}(X,k)=\int d^4 s\, e^{ik\cdot s} 
{\cal G}^{<,\mu\nu}_{\perp}(x,y) \ ,
\end{align}
where $X^\mu=(x^\mu+y^\mu)/2$ and $s^\mu=x^\mu-y^\mu$ are the central and relative coordinate respectively, while $k^\mu$ is the residual momentum. We can build kinetic equations to the desired order in $1/E$ by adding and subtracting the Wigner transformed equations of motion. Precisely, taking into account Eq.\eqref{motion_Green}, we can define
\begin{align}\notag
\left(I_{\pm}\right)^{\mu\nu}=-\dfrac{1}{2}\int d^4 s\, e^{ik\cdot s} \left(\D^2_x \pm \D^2_y \right)\, {\cal G}^{<,\mu\nu}_{\perp}(x,y)= 0  \ .
\end{align}
Performing the Wigner transform, the dispersion $\left(I_{+}\right)^{\mu\nu}=0$ and transport $\left(I_{-}\right)^{\mu\nu}=0$ equations give
\begin{subequations}
\begin{align} \label{dis_OSEFT_1/E^2}
&  \left( 2E(v\cdot k)+k^2-\dfrac{\d^2}{4}\right)\, {\cal G}_{\perp}^{<,\mu\nu}(X,k)= 0
\ ,
\\ \label{tra_OSEFT_1/E^2}
&\left( E \, i v\cdot \d+i k\cdot\d\right) {\cal G}^{<,\mu\nu}_\perp(X,k)=0 \ ,
\end{align}
\end{subequations}
respectively. Above and for the rest of the paper, we will use the shortened notation $ \d^\mu= \d_X^\mu$ when all derivatives in the expressions are respect to the center coordinate. The kinetic equations can also be written in the following form, obtained after solving the theory order by order in $1/E$ (see the remarks in Sec.\ref{II})
\begin{subequations}
\begin{align} 
&  \left(2E(v\cdot k)+k_\perp^2-\dfrac{\d_\perp^2}{4}- \dfrac{(\Tilde{v}\cdot k) k_\perp^2}{2E}+\dfrac{(\Tilde{v}\cdot k)\d_\perp^2}{8E}+\dfrac{(k_\perp\cdot\d)(\Tilde{v}\cdot\d)}{4E} \right)\, {\cal G}_{\perp}^{<,\mu\nu}(X,k)= 0
\ ,
\\ 
&\left( E\, i v\cdot \d+i(k_\perp\cdot\d)-\dfrac{i k_\perp^2 (\Tilde{v}\cdot\d)}{4E}+\dfrac{i(\Tilde{v}\cdot\d)\d_\perp^2}{16E}-\dfrac{i(\Tilde{v}\cdot k) (k_\perp\cdot\d)}{2E}\right) {\cal G}^{<,\mu\nu}_\perp(X,k)=0 \ .
\end{align}
\end{subequations}
In addition, one should complement the kinetic equations with the following constraints which, by construction, are obeyed at any order in $1/E$
\begin{gather}\label{constraints_1}
v_\mu {\cal G}^{<,\mu\nu}_{\perp}=\Tilde{v}_\mu {\cal G}^{<,\mu\nu}_{\perp}=v_\nu {\cal G}^{<,\mu\nu}_{\perp}=\Tilde{v}_\nu {\cal G}^{<,\mu\nu}_{\perp}=0 \ .
\end{gather}
Analogous kinetic equations and constraints can be derived for the negative energy sector, just by replacing $E\rightarrow -E$ and ${\cal G}\rightarrow \widetilde{\cal G}$. The above constraints suggest that the OSEFT Wigner function can be decomposed into the polarization basis introduced in Sec.\ref{II.C} as
\begin{align}\label{WF_pol_basis_pos-general}
{\cal G}^{<,\mu\nu}_{\perp}(X,k)=\sum_{h, h'=\pm}e^{\mu}_{h} e^{*\nu}_{h'}\, {\cal G}^{<,h h'}(X,k)\ .
\end{align}
In the remaining part of the paper we will assume that there is no polarization mixing in the photon ensemble, and thus write
\begin{align}\label{WF_pol_basis_pos}
{\cal G}^{<,\mu\nu}_{\perp}(X,k)=\sum_{h=\pm}e^{\mu}_h e^{*\nu}_h\, {\cal G}^{<,h}(X,k)\ .
\end{align}
The polarization space components are defined as the Wigner transform of the corresponding Green function as
\begin{align}\label{OSEFT_WF_pol_+E}
{\cal G}^{<,h}(X,k)=\int d^4 s \, e^{ik\cdot s} {\cal G}^{<,h}(x,y) \ , 
\end{align}   
where ${\cal G}^{<,h}(x,y)=\braket{\tau_h^\dag(y)\tau_h(x)}$. After the Wigner transform, their general structure is
\begin{align}\label{G_h_pos}
& {\cal G}^{<,h}(X,k)=  2\pi \delta\big(K_{h}\big) f^{h}(X,k) \ ,
\end{align}
where we denote with $f^{h}(X,k)$ the off-shell distribution function for right/left circular polarized photons of positive energy, while $K_{h}$ is the function that governs the dispersion relation, given by the expression inside the parenthesis in Eq.\eqref{dis_OSEFT_1/E^2}. Again, similar definitions can be done for the negative energy sector, e.g $\widetilde{ \cal  G}^{<,h}(X,k)=2\pi \delta\big(\widetilde{K}_{h}\big)\widetilde{f}^{h}(X,k)$. 

%%%%%%%%%%%%%%%%%%%%%%%%%%%%%%%%%%%%%%%%%%%%%%%%%%%%%%%%%%%%
\subsection{Wigner function for photons}
%%%%%%%%%%%%%%%%%%%%%%%%%%%%%%%%%%%%%%%%%%%%%%%%%%%%%%%%%%%%
\label{III.A}

Transport equations associated to the photon Wigner function were studied in the literature long ago, see for example \cite{Vasak:1987um}. The photon Wigner function can be determined in several ways, for instance by using the Fourier decomposition of the vector field gauge potential or by directly solving the quantum kinetic equations \cite{Huang:2020kik,Hattori:2020gqh,Lin:2021mvw}. Here we present yet an alternative derivation using the OSEFT developed in Sec.~\ref{II}. 

The lesser component of the Wigner function for photons is defined as
\begin{equation}\label{full_WF}
G^{<,\mu\nu}(X,q)=\int d^4 s\, e^{i q\cdot s}\braket{A^{\nu}(y)A^\mu(x)}\ .
\end{equation}
Then, plugging the ansatz for the photon field of Eq.\eqref{ansantz_photon_field} onto the above equation and using the momentum decomposition $q^\mu=\pm Ev^\mu+k^\mu$ for the positive/negative energy sector of the theory respectively, one can write
\begin{equation}\label{step1}
G^{<,\mu\nu}(X,q)=\int d^4 s\, e^{i k\cdot s}\left\lbrace\braket{\xi^{\nu\dag} (y)\xi^\mu (x)}+\braket{\xi^{\nu} (y)\xi^{\mu\dag} (x)}\right\rbrace\ ,
\end{equation}
where we used that
\begin{equation}
\braket{\xi^{\nu}(y)\xi^\mu(x)}=\braket{\xi^{\nu\dag}(y)\xi^{\mu\dag}(x)}=0 \ ,
\end{equation}
which is equivalent to impose causality preserving commutation relations between creation and annihilation operators. Then, for free photons we can directly use Eq.\eqref{relation_tau} to write the Wigner function as follows
\begin{equation}
\begin{gathered} \label{full_WF_OSEFT} 
G^{<,\mu\nu}(X,q)=\int d^4s \, e^{ik\cdot s} \bigg\lbrace  \left(\mathcal{O}^{\mu}_{\ \alpha}\right)_x
\left(\mathcal{O}^{\nu}_{\ \beta}\right)_y^*
{\cal G}^{<,\alpha\beta}_{\perp}(x,y) +\left(\mathcal{O}^{\mu}_{\ \alpha}\right)_x^*
\left(\mathcal{O}^{\nu}_{\ \beta}\right)_y
\widetilde{\cal G}^{<,\alpha\beta}_{\perp}(x,y)\bigg\rbrace
 \ , 
\end{gathered}
\end{equation}
where ${\cal G}^{<,\alpha\beta}_{\perp}(x,y)$ and $\widetilde{\cal G}^{<,\alpha\beta}_{\perp}(x,y)$ are the OSEFT Green functions for positive and negative energies introduced in Eqs.(\ref{RT_Green_pos}-\ref{RT_Green_neg}) respectively, and we used the shortened notation
\begin{equation}
\mathcal{O}^{\mu}_{\ \nu}= g^\mu_\nu-n^\mu\d_\nu\frac{ n\cdot\D}{\D^2+( n\cdot\D)^2} \ .   
\end{equation}
Note that in Eq.\eqref{full_WF_OSEFT}, we assumed that
\begin{equation}
\braket{\phi ^\dag(y)\tau_{\perp}^{\mu}(x)}=\braket{\tau_{\perp}^{\mu\dag}(y)\phi (x)}=\braket{\phi ^\dag(y)\phi (x)}=0 \ ,
\end{equation}
which can be justified by noting that in the Lagrangian of Eq.\eqref{Lagrangian_OSEFT} there are no terms that couple those fields. Then, using the expansion $\mathcal{O}^{\mu}_{\ \nu} \approx g^\mu_{\nu}-\frac{i}{E} n^\mu\d_\nu$ and decomposing the OSEFT Wigner functions onto the circular polarization basis, we reach to
\begin{equation}
\begin{gathered}
G^{<,\mu\nu}(X,q)=\sum_{h={\pm}} \int d^4s \, e^{ik\cdot s}\bigg\lbrace \left(e^{\mu}_h e_h^{*\nu}+\dfrac{i}{E}\left(e^{\mu}_h n^\nu(e_h^*\cdot\d_y)-n^\mu e_h^{*\nu}(e_h\cdot\d_x)\right)\right)
{\cal G}^{<,h}(x,y) 
\\
+\left(e^{*\mu}_h e_h^{\nu}-\dfrac{i}{E}\left(e^{*\mu}_h n^\nu(e_h\cdot\d_y)-n^\mu e_h^\nu(e_h^*\cdot\d_x)\right)\right)
\widetilde{\cal G}^{<,h}(x,y)
\bigg\rbrace\ .
\end{gathered}   
\end{equation}
By writing $\d_x^\mu=\frac{1}{2}\d_X^\mu+\d_s^\mu$ and $\d_y^\mu=\frac{1}{2}\d_X^\mu-\d_s^\mu$ above, we can easily perform the Wigner transform, as $\d_s^\mu\rightarrow -ik^\mu$ after integrating by parts. The result may be written as
\begin{align} \label{G_FULL_Pi}
G^{<,\mu\nu}(X,q)=\sum_{h=\pm} \left(\Pi^{\mu\nu}_{h}(k) {\cal G}^{<,h}(X,k)+\widetilde{\Pi}^{\mu\nu}_{h}(k)\widetilde{\cal G}^{<,h}(X,k)\right)\ ,
\end{align}   
where
%now ${\cal G}^{<,h}(X,k)$ and $\widetilde{G}^{<,h}(X,k)$ are the polarization space components of the OSEFT Wigner functions 
 we defined the tensor
\begin{align}\label{Pi_k}
\Pi^{\mu\nu}_{h}(k)=e^{\mu}_h e^{*\nu}_h -\dfrac{1}{E}\left(e^{\mu}_h n^\nu (e_h^*\cdot k)+n^\mu e^{*\nu}_h (e_h\cdot k) \right)+\dfrac{i}{2E}\left(e^{\mu}_h n^\nu (e_h^*\cdot \d)-n^\mu e^{*\nu}_h (e_h\cdot \d)  \right)\ ,
\end{align}
while $\widetilde{\Pi}^{\mu\nu}_h(k)$ can be obtained after replacing $E\rightarrow -E$ and $h\rightarrow -h$ in the above expression. Note that in the above tensor we are reproducing the expansion of the full theory polarization vectors of Eq.\eqref{P_h_pol_vectors}. The tensor $\Pi_h^{\mu\nu}(k)$ can also be written in terms of the transverse projector and the spin tensor
\begin{align}\label{Pi_k_tensors}
\Pi^{\mu\nu}_{h}(k)=-\dfrac{1}{2}\left( P_{\perp}^{\mu\nu}+hS_\perp^{\mu\nu} -\dfrac{1}{E}n^{(\mu} k^{\nu)}_\perp +\dfrac{h}{E} n^{[\mu} S_{\perp}^{\nu]\alpha}  k_{\alpha}-\dfrac{i}{2E} n^{[\mu} \d^{\nu]}_\perp+\dfrac{ih}{2E}n^{(\mu} S_{\perp}^{\nu)\alpha}\d_{\alpha}
\right)\ ,
\end{align}
after resorting to Eq.\eqref{P_h}. Above, it can be easily seen that we are reproducing the expansions of the full theory transverse projector and spin tensor of Eqs.(\ref{P_perp_full_EFT}-\ref{S_perp_full_EFT}) up to first order in $1/E$, additionally, we produce quantum corrections to the Wigner function. 

Now we would like to re-express the Wigner function in terms of the full theory variables. At this expansion order we can associate
\begin{subequations}
\begin{align}\label{rule_1}
&  {\cal G}^{<,h}(X,k)=2\pi \delta(K_{h}) f^{h}(X,k)\, \longrightarrow \, 4\pi \delta(q^2)\theta(u\cdot q)f^{h}(X,q) \ ,
\\ \label{rule_2}
&  \widetilde{\cal G}^{<,h}(X,k)=2\pi \delta(\widetilde{K}_{h}) \widetilde{f}^{h}(X,k)\, \longrightarrow \, -4\pi \delta(q^2)\theta(-u\cdot q)f^{h}(X,q) \ ,
\end{align}
\end{subequations}
where $f^{h}(X,q)$ is the off-shell distribution for right/left handed circularly polarized photons of the full theory. If, for instance, one considers that the photon ensemble is at thermal equilibrium and that there is no CP violating effect, so that one can assume $f^{+}(X,q)=f^{-}(X,q)\equiv f_{\text{eq}}(q_0)$, the distribution function takes the form $f_{\text{eq}}(q_0)=(e^{q_0/T}-1)^{-1}$ in the rest frame of the medium, being $T$ the equilibrium temperature. In that scenario, one has
\begin{equation}
f_{\text{eq}}(q_0)=
\begin{cases}
n_B(\q) \ , & \text{if} \quad q_0=\vert\q\vert  
\\
-[1+n_B(\q)] \ , & \text{if} \quad q_0=-\vert\q\vert 
\end{cases}
\ ,
\end{equation}
where $n_B(\q)=(e^{\vert\q\vert/T}-1)^{-1}$ is the Bose-Einstein distribution function. The tensors $\Pi^{\mu\nu}_h(k)$ and $\widetilde{\Pi}^{\mu\nu}_h(k)$ both translate to $\Pi^{\mu\nu}_h(q)$, which may be written either in terms of the polarization vectors or the transverse projector and spin tensor as (see Sec.(\ref{II.F}))
\begin{subequations}
\begin{align} \label{WF_1/E_pol}
& \Pi^{\mu\nu}_h(q)=e^{\mu}_{q,h} e^{*\nu}_{q,h}+\dfrac{i}{2E_q}\left(e^{\mu}_{q,h}n^\nu_q (e_{q,h}^*\cdot \d)-n^\mu_q e^{*\nu}_{q,h} (e_{q,h}\cdot \d) \right) \ ,
\\ \label{WF_1/E_tensor}
& \Pi^{\mu\nu}_h(q)=-\dfrac{1}{2}\left(P_{\perp,q}^{\mu\nu}+hS_{\perp,q}^{\mu\nu}
-\dfrac{i}{2E_q}n_q^{[\mu} \d^{\nu]}_{\perp,q}+\dfrac{ih}{2E_q}n_q^{(\mu} S_{\perp,q}^{\nu)\alpha} \d_{\alpha} \right)\ ,
\end{align}
\end{subequations}
where we recall that $n^\mu_q=u^\mu-v^\mu_q$, being $v^\mu_q=q^\mu/E_q$ the photon on-shell velocity. Then, after moving back to the full theory variables, one can write the Wigner function for photons at $1/E_q$ accuracy as
\begin{equation}\label{WF_1/E}
 G^{<,\mu\nu}(X,q)=4\pi\sum_{h=\pm}  \Pi^{\mu\nu}_h(q)
\delta(q^2)\text{sgn}(u\cdot q)f^{h}(X,q) \ ,
\end{equation}
with $\text{sgn}(x)$ denoting the sign function. The kinetic equations obeyed by the photon Wigner function can be deduced from Eqs.(\ref{dis_OSEFT_1/E^2}-\ref{tra_OSEFT_1/E^2}). In terms of the full theory variables they read
\begin{subequations}
\begin{align} \label{dispersion-full}
&  q^2\, G^{<,\mu\nu}(X,q)= 0
\ ,
\\ 
\label{transport-full}
&\left( q\cdot \d\right) G^{<,\mu\nu}(X,q)=0 \ .
\end{align}
\end{subequations}
In Eq.~(\ref{dispersion-full}) we dropped a piece proportional 
to  $\sim\d^2$ to be consistent with the gradient expansion assumed here.
We note that at this expansion order the following constraints are satisfied
\begin{equation}
\left(\dfrac{1}{2}\d_\mu-iq_\mu\right)\Pi^{\mu\nu}_h(q)= \left(\dfrac{1}{2}\d_\nu+iq_\nu\right)\Pi^{\mu\nu}_h(q)=0 \ ,
\end{equation}
and also 
\begin{equation}
u_\mu \Pi^{\mu\nu}_h(q)  = u_\nu\Pi^{\mu\nu}_h(q)=0 \ ,
\end{equation}
so that the Wigner function obeys the Coulomb gauge-fixing conditions. We note that the Wigner function of Eq.\eqref{WF_1/E} coincides exactly with that encountered in Refs.\cite{Huang:2020kik,Hattori:2020gqh,Lin:2021mvw}. However, let us remark that we did not impose any gauge fixing condition to achieve this result, which means that the Wigner function associated to the high-energy modes, when computed in a semi-classical approach, is a gauge invariant quantity, provided that the gauge transformations are respectful with the separation of energy scales

%%%%%%%%%%%%%%%%%%%%%%%%%%%%%%%%%%%%%%%%%%%%
\subsection{Side jumps from reparametrization invariance}
%%%%%%%%%%%%%%%%%%%%%%%%%%%%%%%%%%%%%%%%%%%%
\label{III.B}

In a semi-classical approach to describe massless chiral fermions
the Lorentz transformations need to be modified in the presence of small quantum effects  in order to preserve the frame independence of the theory \cite{Chen:2014cla}. This is related to the fact that the total angular momentum of a relativistic spinning particle is ambiguous, because the definition of the spin part is non-unique. The issue is resolved by imposing the condition $u_\mu S_{\perp,q}^{\mu\nu}=q_\mu S_{\perp,q}^{\mu\nu}=0$, so that the spin tensor is uniquely fixed in the inertial frame, and is given then by Eq.\eqref{S_perp_full}. Consequently, when moving from frame $u^\mu$ to $u^{\prime\mu}$ the particle position also changes, so that the total angular momentum is conserved. The shift on the particle position when changing between inertial frames is the so called side jump effect \cite{Chen:2014cla,Chen:2015gta}. In the context of  chiral kinetic theory,  describing a system of massless  fermions, this effect is manifested by the fact that the fermion distribution function is no longer a Lorentz scalar.

In chiral kinetic theory, the side jump effect can naturally be derived from OSEFT, and it is linked with
the reparametrization invariance of the theory. 
This was checked for massless fermions in  \cite{Carignano:2018gqt,Carignano:2019zsh}. As we will see, the same applies to photons. Note that while it has been widely accepted that this side jump effect would affect also other massless spinning particles \cite{Huang:2018aly}, and not only fermions, we are however unaware of any explicit proof of this side jump effect from quantum field theory.

In order to derive the side jump effect, we first need to know the RI transformations of the
polarization space components of the Wigner function. Those can be derived from 
\begin{equation}
{\cal G}^{<,h}(X,k)= e_{h,\mu}^* e_{h,\nu}\int d^4 s \, e^{ik\cdot s} \braket{\tau_\perp^{\dag\nu }(y)\tau_{\perp}^{\mu}(x)} \ ,
\end{equation}
by employing the transformation rules of Eqs.(\ref{RI_I_tau_LFR_1/E}-\ref{RI_III_tau_LFR_1/E}). Under a type (II) transformation one finds
\begin{align} \notag
& \delta_{(\text{II})} {\cal G}^{<,h}(X,k)  = \dfrac{i}{2E} e_{h,\mu}^* e_{h,\nu} \int d^4 s\, e^{ik\cdot s} \bigg( (\d_{x}+\d_y)^\mu \Tilde{\Delta}_\perp^\nu-\Tilde{\Delta}_\perp^\mu(\d_{x}+\d_y)^\nu \bigg) {\cal G}^{<,h}(x,y)=
\\ 
&\qquad \qquad \qquad \  =\dfrac{ih}{2E}S_\perp^{\mu\nu}\Tilde{\Delta}_{\perp,\nu} \d_{\mu} {\cal G}^{<,h}(X,k) \ .
\end{align}    
The contribution from the negative energy sector, $\delta_{(\text{II})}\widetilde{\cal G}^{<,h}(X,k)$, is the same as above, as can be seen by replacing $E\rightarrow -E$ and $h\rightarrow -h$. After adding the positive and negative energy sector contribution, we can deduce the transformation rule for the photon distribution function under a type (II) transformation, which reads
\begin{align}\label{Side_jump}
f^h(X,q)\overset{\text{(II)}}{\longrightarrow } f^h(X,q)+\dfrac{ih}{2E_q}S_{\perp,q}^{\mu\nu}\Tilde{\Delta}_{\perp,\nu}\d_{\mu} f^{h}(X,q) \ ,
\end{align}
after moving back to the full theory variables, which is the expected side jump effect. Note that
Eq.~(\ref{Side_jump}) gives the infinitesimal change of the distribution function, where one has to take into account that $\Tilde{\Delta}_{\perp}^\mu /2 =u^{\prime\mu}-u^\mu$.

Doing a similar reasoning as above, one can show that at $1/E_q$ accuracy $S^{\mu\nu}_{\perp,q}$ and $f^h(X,q)$ are both invariant under a type (I) and a type (III) RI transformation, so that there is no side jump effect in those cases.

%%%%%%%%%%%%%%%%%%%%%%%%%%%%%%%%%
\section{Stokes parameters with quantum corrections}
%%%%%%%%%%%%%%%%%%%%%%%%%%%%%%%%
\label{IV}

By projecting the Wigner function of Eq.\eqref{WF_1/E} onto the circular polarization basis, we can build a naive photon current associated to every helicity state. Precisely, in the absence of polarization mixing in the photon ensemble, we can define
\begin{equation}\label{current}
j^{h,\mu}(X)=\int \dfrac{d^4q}{(2\pi)^4} q^\mu G^{h}(X,q) \ , \quad h=\pm \ ,
\end{equation}
where we dropped the lesser symbol $(<)$ from the Wigner function to enlighten the notation. Also, we introduced 
\begin{equation}
G^{h}(X,q)=4\pi
\delta(q^2)\text{sgn}(u\cdot q)f^{h}(X,q) \ .
\end{equation}
The polarization space components of the Wigner function (or the zeroth component of the photon current), can be directly related to the Stokes parameters \cite{jackson,Beneke:2010eg}. Then, with this setting, the naive Stokes parameters matrix can be defined as
\begin{equation}
\rho(X)=
\begin{pmatrix}
j^{+}_0(X) & 0 \\
0 &  j^{-}_0(X)
\end{pmatrix}
=
\begin{pmatrix}
j^{I}_0(X)-j^{V}_0(X) & 0 \\
0 & j^{I}_0(X)+j^{V}_0(X) 
\end{pmatrix}
\ ,
\end{equation}
where $j^{I}_0= ( j_0^+ + j_0^-)/2 $ and $j^{V}_0= -( j_0^+ - j_0^-)/2 $ give the values of the intensity and degree of polarization of the photon ensemble, respectively. It is well-known that the Stokes parameters are not Lorentz invariant, but the percentage of circulation polarization, obtained as the ratio of the degree of circular polarization over the intensity, i.e $j^{V}_0/j^{I}_0$,
is usually regarded in the literature as a Lorentz invariant \cite{Cocke,Han:1997zz}.  

However, in a semi-classical approach to photon propagation based on quantum kinetic theory, as soon as quantum corrections are considered, and due to the side jump effect discussed in Sec. \ref{III.B} this is no longer the case. Modification of the definition of the Stokes parameters is then needed in order to restore the Lorentz invariance of the polarization ratios. 

The above issue is related to the fact that the naive current of Eq.\eqref{current} does not transform as a Lorentz vector when quantum corrections are taken into account, because the photon distribution function is no longer a Lorentz scalar. A solution was found in Ref.\cite{Chen:2014cla}, by including a magnetization contribution to the naive  helicity current one can restore its frame independence in the collisionless limit. Indeed, one can define a frame independent photon current in the collisionless limit as
\begin{equation}\label{current}
J^{h,\mu}(X)= \int \dfrac{d^4q}{(2\pi)^4} \left(q^\mu -ih S^{\mu\nu}_{\perp,q}\d_\nu \right)G^{h}(X,q) \ .
\end{equation}
From the above current, we can define new Stokes parameters as $J^{I}_0= ( J_0^+ + J_0^-)/2 $ and $J^{V}_0= -( J_0^+ - J_0^-)/2 $, so that the Lorentz invariance of the polarization ratio $J^V_0/J_0^I$ is restored. Hence, we can write the intensity and the degree of polarization of the photon ensemble as
\begin{align} \label{JI}
& J^{I}_0(X)= \int \dfrac{d^4q}{(2\pi)^3} 2\delta(q^2)\text{sgn}(u\cdot q)\left( q_0 f^{I}(X,q) + \dfrac{ \u\times \q}{u\cdot q}\cdot \boldsymbol{\nabla}f^V(X,q)\right) \ ,
\\ \label{JV}
& J^{V}_0(X)= \int \dfrac{d^4q}{(2\pi)^3} 2\delta(q^2)\text{sgn}(u\cdot q)\left( q_0f^{V}(X,q) +\dfrac{\u\times \q}{u\cdot q}\cdot \boldsymbol{\nabla}  f^I(X,q)\right) \ ,
\end{align}
respectively, where the we have defined
\begin{align}\label{fI}
& f^{I}(X,q)=\dfrac{f^{+}(X,q)+f^{-}(X,q)}{2} \ ,
\\ \label{fV}
& f^{V}(X,q)=-\dfrac{f^{+}(X,q)-f^{-}(X,q)}{2} \ .
\end{align}
Note that in the rest frame of the medium ($\boldsymbol{u}=\boldsymbol{0}$) or in the frames satisfying $\u\times \q=0$, the naive polarization percentage $j^{V}_0/j^{I}_0$ is still a LI quantity, even in the presence of small quantum effects.

Let us emphasize that the frame independence of the current of Eq.\eqref{current} is only valid under the assumption of a collisionless medium. Nevertheless, under certain circumstances this program can also be generalized in the presence of collisions \cite{Chen:2015gta}, so that further modification of the Stokes parameters would be required in that scenario.

%%%%%%%%%%%%%%%%%%%%%%%%%%%%%%%%%%%%%%%%%%%%
\section{Discussion}
%%%%%%%%%%%%%%%%%%%%%%%%%%%%%%%%%%%%%%%%%%%%
\label{V}

In this manuscript we have fully developed the OSEFT associated to photons, generalizing  previous work carried out for chiral fermions. Assuming a high energy expansion, we derived the Lagrangian associated to the high energy or hard photons from the free Maxwell Lagrangian. 
By splitting the vector gauge field into different components, we identified the  degree of freedom that can be integrated out from the Lagrangian using its classical equation of motion, given in Eq.\eqref{motion_lambda}. Subsequently, we showed that the remaining non-physical degree of freedom can be eliminated employing the local field redefinition of Eq.\eqref{LFR_all_orders}. After this last step, we obtained an effective field theory that only contains a transverse vector gauge field  (cf. Eq. \eqref{Lagrangian_OSEFT}). 

Generally, in the context of effective field theories, it is common to split the fields into different components, according to their energy scale (e.g into hard and soft parts). In a gauge theory then one expects a gauge symmetry associated to each sector of the theory which is respectful with the energy scaling of such decomposition. In this work, a gauge symmetry associated to the hard part of the photon field has been presented, and we demonstrated that the OSEFT Lagrangian enjoys that symmetry.  We have also proven the RI of the theory, which basically means that the OSEFT is respectful with the Lorentz symmetry. Using the RI transformations, a first principles derivation of the so called side jump effect for photons has been provided (see Eq.\eqref{Side_jump}).

Since at high energies, or when the photon wavelength is much shorter than any length scale in the system, electromagnetic waves can be accurately studied under the so called eikonal limit, the OSEFT seems a suitable tool to study how the physics beyond the eikonal approximation is corrected. For instance, we used the OSEFT and the Schwinger-Keldysh formalism of thermal field theory to construct a quantum kinetic theory for photons. We have shown how to use the OSEFT to systematically compute quantum corrections to the semi-classical photon Wigner function and we explicitly computed the leading order quantum correction (see Eqs.\ref{WF_1/E_pol}-\ref{WF_1/E_tensor} and Eq.\eqref{WF_1/E}). In addition, we derived the quantum kinetic equations obeyed by the photon Wigner function in the collisionless limit, given by Eqs.(\ref{dispersion-full}-\ref{transport-full}).                                             Our results agree with those found in Refs.\cite{Huang:2020kik,Hattori:2020gqh,Lin:2021mvw}. However, as in the effective field theory approach we separate the gauge field into physical and non-physical components, in OSEFT computations one can clearly identify which modes contribute at each order in the energy expansion, thus providing valuable insight.

We have then elaborated on the proper definition of the Stokes parameters as deduced from quantum kinetic theory. At the classical level, the percentage of circular polarization in a system is a Lorentz invariant quantity. However, we have shown that as soon as quantum corrections are considered, the definition of the Stokes parameters needs to be modified in order to preserve the Lorentz invariance of the percentage of circular polarization. We have given such a definition in Eqs.(\ref{JI}-\ref{JV}) for the intensity and degree of circular polarization. Let us recall that we assumed the absence of polarization mixing in the photon ensemble. It would be interesting to generalize the discussion in the presence of such polarization modes, as they are known to be generated in relevant cosmological physical scenarios. 

Finally, 
let us mention that the OSEFT for photons could be used also for the computation of power corrections to different sort of Feynman diagrams, for example, at high temperature and/or density.
One could extend the study we carry out by allowing interactions of the hard photons with either hard and soft fermions, and proceed with the same methods for the study of the effects of these interactions, as it was done for the contributiuon of hard fermions in \cite{Manuel:2016wqs,Manuel:2016cit}.

{\bf Acknowledgements} 
	
We thank Joan Soto and Stefano Carignano for  discussions.
This work was supported by Ministerio de Ciencia, Investigaci\'on y Universidades (Spain) MCIN/AEI/10.13039/501100011033/ FEDER, UE, under the project PID2022-139427NB-I00, by Generalitat de Catalunya by the project 2021-SGR-171 (Catalonia). This work was also partly supported by the Spanish program Unidad de Excelencia
 Maria de Maeztu CEX2020-001058-M.
 %financed by MCIN/AEI/10.13039/501100011033.

\appendix

%%%%%%%%%%%%%%%%%%%%%%%%%%%%%%%%%%%%%%%%%%%%%%%%%%%
\section{Operator notation and identities}
\label{A}
%%%%%%%%%%%%%%%%%%%%%%%%%%%%%%%%%%%%%%%%%%%%%%%%%%%

In this manuscript we use a simplified operator notation in order to shorten the length of the expressions. When we write one over a differential operator
 we mean the inverse of the differential operator in the denominator. 
To compactify even more the notation, we often write fractions of different differential operators. For example it should be understood the following
\begin{equation}\label{motion_lambda_simplified}
%\frac{1}{\D^2+(n\cdot\D)^2}(n\cdot\D)\partial_\mu \xi_{\perp}^{\mu}(x)\longrightarrow  
\frac{n\cdot\D}{\D^2+(n\cdot\D)^2}\partial_\mu \xi_{\perp}^{\mu}
\longrightarrow 
\frac{1}{\D^2+(n\cdot\D)^2}(n\cdot\D)\partial_\mu \xi_{\perp}^{\mu}
\ .
\end{equation}
%

%When we find expressions as 
%
%\begin{equation}%\label{motion_lambda_simplified}
%\frac{1}{\D^2+(n\cdot\D)^2}(n\cdot\D)\partial_\mu \xi_{\perp}^{\mu}(x)\longrightarrow  \frac{n\cdot\D}{\D^2+(n\cdot\D)^2}\partial_\mu \xi_{\perp}^{\mu}(x) \ ,
%\end{equation}
%
%in the equation of motion of Eq.\eqref{motion_lambda}. 

Also, in most of the derivations presented in this work, we used the fact that some combinations of operators, which appear acting on the $\phi(x)$ field, vanish identically. For instance, when deriving Eq.\eqref{relation_tau} we used the fact that
\begin{equation}\label{operator_identity}
\dfrac{n\cdot\D}{\D^2+(n\cdot\D)^2}\dfrac{\d_\perp^2}{u\cdot\D}-\dfrac{\D^2-(v\cdot\d)(n\cdot\D)}{\D^2+(n\cdot\D)^2}+\dfrac{v\cdot\d}{u\cdot\D}=0 \ .    
\end{equation}
Similarly, one can show other operator identities which are relevant in our derivations
\begin{subequations}
\begin{align}
& \dfrac{u\cdot\D}{\D^2+(n\cdot\D)^2}\dfrac{\D^2}{n\cdot\D}-\dfrac{\D^2-(v\cdot\d)(n\cdot\D)}{\D^2+(n\cdot\D)^2}-\dfrac{v\cdot\d}{n\cdot\D}=0 \ ,
\\
& \dfrac{(u\cdot\D)(n\cdot\D)(\D^2-(v\cdot\d)(n\cdot\D))}{\D^2+(n\cdot\D)^2}-(\D^2-(v\cdot\d)(u\cdot\D))+\dfrac{\D^2\d_\perp^2}{\D^2+(n\cdot\D)^2}=0 \ .
\end{align}
\end{subequations}
%

%%%%%%%%%%%%%%%%%%%%%%%%%%%%%%%%%%%%%%%%%%%%%%%%%%%
\section{RI transformations}
%%%%%%%%%%%%%%%%%%%%%%%%%%%%%%%%%%%%%%%%%%%%%%%%%%%
\label{B}

As we explained in Sec.(\ref{II.E}), the transformation rules are modified after integrating out the $\lambda(x)$ field. Precisely, for the transverse field we find
\begin{align}\label{RI_tau_I}
& \xi^{\mu}_{\perp} \overset{\text{(I)} }{\longrightarrow } \xi^{\mu}_{\perp}-\dfrac{\Delta_\perp^\mu}{2}\left(1+\dfrac{(u\cdot\D)(n\cdot\D)}{\D^2+(n\cdot\D)^2}\right)\phi-\dfrac{\Delta_\perp^\mu}{2}\dfrac{(n\cdot\D)(\d\cdot\xi_\perp)}{\D^2+(n\cdot\D)^2}-\frac{v^\mu}{2}(\Delta_\perp\cdot\xi) -n^\mu \delta_{(\text{I})} \lambda \ ,
\\ \label{RI_tau_II}
& \xi^{\mu}_{\perp} \overset{\text{(II)}  }{\longrightarrow } \xi^{\mu}_{\perp}+\dfrac{\Tilde{\Delta}_{\perp}^\mu}{2}\left(\frac{ (n\cdot \D)(\d\cdot \xi_\perp)}{\D^2+(n\cdot\D)^2} -\frac{\D^2-(v\cdot\d)(n\cdot \D)}{\D^2+(n\cdot\D)^2}\phi\right)-\frac{v^\mu}{2}(\Tilde{\Delta}_\perp\cdot \xi) -n^\mu \delta_{(\text{II})} \lambda \ ,
\\ \label{RI_tau_III}
& \xi^{\mu}_{\perp} \overset{\text{(III)}\,}{\longrightarrow } \xi^{\mu}_{\perp}-\alpha\, n^\mu\left(\frac{( n\cdot \D)(\partial\cdot \xi_{\perp})}{\D^2+(n\cdot\D)^2} -\frac{\D^2-(v\cdot\d)(n\cdot \D)}{\D^2+(n\cdot\D)^2}\phi\right) -n^\mu\delta_{(\text{III})} \lambda \ .
\end{align}   
In the above transformations, we defined the quantities $\delta_{(\Lambda)} \lambda$ for $\Lambda=\lbrace\text{I, II, III}\rbrace$ given by 
\begin{align}\notag
& \delta_{(\text{I})}\lambda= \dfrac{1}{2} \dfrac{\D^2+2(n\cdot\D)^2}{\D^2+(n\cdot\D)^2}(\Delta_\perp\cdot \xi_\perp)+\dfrac{1}{2} \dfrac{\D^2}{(\D^2+(n\cdot\D)^2)^2}(\Delta_\perp\cdot \d_\perp)(\d\cdot\xi_\perp)
\\\label{deltalambda_I}
&\ \qquad +\dfrac{1}{2}\left(\dfrac{(n\cdot\D)(\D^2-(v\cdot\d)(n\cdot\D))}{(\D^2+(n\cdot\D)^2)^2}+\dfrac{v\cdot\d}{\D^2+(n\cdot\D)^2}\right)(\Delta_\perp\cdot\d)\phi \ ,
\\\notag
& \delta_{(\text{II})}\lambda= \dfrac{1}{2} \dfrac{\D^2}{\D^2+(n\cdot\D)^2}(\Tilde{\Delta}_\perp\cdot \xi_\perp)-\dfrac{1}{2} \dfrac{\D^2}{(\D^2+(n\cdot\D)^2)^2}(\Tilde{\Delta}_\perp\cdot \d_\perp)(\d\cdot\xi_\perp)
\\ \label{deltalambda_II}
& \ \qquad-\dfrac{1}{2}\left(\dfrac{(n\cdot\D)(\D^2-(v\cdot\d)(n\cdot\D))}{(\D^2+(n\cdot\D)^2)^2}+\dfrac{v\cdot\d}{\D^2+(n\cdot\D)^2}\right)(\Tilde{\Delta}_\perp\cdot\d)\phi \ ,
\\\notag
& \delta_{(\text{III})}\lambda= -\alpha\left(\frac{ n\cdot \D}{\D^2+(n\cdot\D)^2} (\d\cdot\xi_\perp)-\frac{\D^2-(v\cdot\d)(n\cdot \D)}{\D^2+(n\cdot\D)^2}\phi \right) +\alpha\dfrac{\D^2(u\cdot \D)}{(\D^2+(n\cdot\D)^2)^2}(\d\cdot\xi_\perp)
\\ \label{deltalambda_III}
& \ \qquad+\alpha\left(\dfrac{(u\cdot\D)(n\cdot\D)(\D^2-(v\cdot\d)(n\cdot\D))}{(\D^2+(n\cdot\D)^2)^2}-\dfrac{\D^2-(v\cdot\d)(u\cdot\D)}{\D^2+(n\cdot\D)^2}\right)\phi \ , 
\end{align}    
which account for the infinitesimal variations of the operators on the equation of motion of $\lambda(x)$ (c.f Eq.\eqref{motion_lambda}). For the $\phi(x)$ field, only the transformation rule under type (III) needs to be modified
\begin{equation} \label{RI_phi_II}
 \phi  \overset{\text{(III)}\, }{\longrightarrow }  (1-\alpha)\phi-\alpha \left(\frac{ n\cdot \D}{\D^2+(n\cdot\D)^2} (\d\cdot \xi_\perp)-\frac{\D^2-(v\cdot\d)(n\cdot \D)}{\D^2+(n\cdot\D)^2}\phi\right)\ .
\end{equation}
Remarkably, using Eqs.(\ref{RI_tau_I}-\ref{RI_tau_III}) and the corresponding transformations for the $\phi(x)$ field, one can show that the Lagrangian of Eq.\eqref{L_integrated}, obtained after integrating out the hard field $\lambda(x)$, is RI invariant. Finally, let us conclude by writing down the general transformations for the locally redefined field of Eq.\eqref{LFR_all_orders}, they read
\begin{align} \notag 
\tau_{\perp}^{\mu} \overset{\text{(I)}}{\longrightarrow } \tau_{\perp}^{\mu}-\dfrac{\d_\perp^\mu}{2(u\cdot \D)}(\Delta_\perp\cdot\tau_\perp  )-\dfrac{\Delta_\perp^\mu}{2}\dfrac{n\cdot\D}{\D^2+(n\cdot\D)^2}(\d\cdot\tau_\perp  )-\dfrac{v^\mu}{2}(\Delta_\perp\cdot \tau_\perp) &
\\  \label{RI_type_I_final}
-\dfrac{n^\mu}{2}\bigg( \dfrac{\D^2+2(n\cdot\D)^2}{\D^2+(n\cdot\D)^2}(\Delta_{\perp}\cdot\tau_\perp  )+\dfrac{\D^2}{(\D^2+(n\cdot\D)^2)^2}(\Delta_{\perp}\cdot\d)(\d\cdot\tau_\perp  )\bigg)\ , &
\\ \notag
\tau_{\perp}^{\mu} \overset{\text{(II)}}{\longrightarrow } \tau_{\perp}^{\mu}-\dfrac{\d_\perp^\mu}{2(u\cdot \D)}(\Tilde{\Delta}_\perp\cdot\tau_\perp  )+\dfrac{\Tilde{\Delta}_\perp^\mu}{2}\dfrac{n\cdot\D}{\D^2+(n\cdot\D)^2}(\d\cdot\tau_\perp  )-\frac{v^\mu}{2}(\Tilde{\Delta}_\perp\cdot \tau_\perp  ) \ , &
\\ \label{RI_type_II_final}
-\dfrac{n^\mu}{2}\bigg( \dfrac{\D^2}{\D^2+(n\cdot\D)^2}(\Tilde{\Delta}_{\perp}\cdot\tau_\perp  )-\dfrac{\D^2}{(\D^2+(n\cdot\D)^2)^2}(\Tilde{\Delta}_{\perp}\cdot\d)(\d\cdot\tau_\perp  )\bigg)\ , &
\\ \label{RI_type_III_final}
\tau_{\perp}^{\mu} \overset{\text{(III)}}{\longrightarrow } \tau_{\perp}^{\mu}+\alpha\dfrac{\d_\perp^\mu}{u\cdot\D}\dfrac{n\cdot\D}{\D^2+(n\cdot\D)^2} (\d\cdot\tau_\perp  )+\alpha \, n^\mu\dfrac{\D^2(u\cdot\D)}{(\D^2+(n\cdot\D)^2)^2} (\d\cdot\tau_\perp  )\ .&
\end{align}   

\end{document}